\documentclass[journal]{IEEEtran}
\usepackage{amsmath,latexsym,amssymb,stmaryrd,pifont}
\usepackage{setspace,stackrel,tikz,graphicx}
\usepackage{exscale,relsize,subfig,textcomp,stackrel,setspace,float}
\usepackage{mdframed,pifont}
\usepackage{multicol,xfrac}
\usepackage{multirow, booktabs}
\usepackage{mathtools}
\usepackage{tabularx}
\usepackage[input-decimal-markers=.]{siunitx}
\usepackage{longtable}
\usepackage[T1]{fontenc}
\usepackage{fancyhdr}
\usepackage[font=footnotesize]{caption}
\usepackage{lipsum}
\usepackage{multicol}
\usepackage{graphicx}
\usepackage[utf8]{inputenc}
\pdfoutput=1
\usepackage{amsthm}
\usepackage{mathtools,amssymb,lipsum}

\usepackage{cuted}
\begin{document}

\title{\vspace{0cm}\LARGE Layered Chirp Spread Spectrum Modulations for LPWANs\vspace{0em}}
\makeatletter
\patchcmd{\@maketitle}
  {\addvspace{0\baselineskip}\egroup}
  {\addvspace{0\baselineskip}\egroup}
  {}
  {}
\makeatother
\author{Ali Waqar Azim, Ahmad Bazzi, Roberto Bomfin, Raed Shubair, Marwa Chafii
\thanks{Ali Waqar Azim is with Department of Telecommunication Engineering,  University of Engineering and Technology,  Taxila,  Pakistan (email: aliwaqarazim@gmail.com).}
\thanks{Ahmad Bazzi, Roberto Bomfin and Raed Shubair and Marwa Chafii are with the Engineering Division, New York University (NYU) Abu Dhabi, 129188, UAE
(email: {ahmad.bazzi,raed.shubair, marwa.chafii}@nyu.edu). Marwa Chafii is also with NYU WIRELESS, NYU Tandon School of Engineering, Brooklyn, 11201, NY, USA .}}
\maketitle
 \thispagestyle{fancy}
 \lhead{A. W. Azim, A. Bazzi, R. Bomfin, R. Shubair, M. Chafii, ``Layered Chirp Spread Spectrum Modulations for LPWANs,'' \textit{in IEEE Transactions on Communications}, Vol. 72, Issue 3, March 2024.}
\renewcommand{\headrulewidth}{0pt}

\begin{abstract}
This article examines two chirp spread spectrum techniques specifically devised for low-power wide-area networks (LPWANs) to optimize energy and spectral efficiency (SE). These methods referred to as layered CSS (LCSS) and layered dual-mode CSS (LDMCSS), involves utilizing multiple layers for multiplexing symbols with varying chirp rates. These waveform designs exemplify a high degree of SE compared to existing schemes. Additionally, LDMCSS necessitates a lesser number of layers than LCSS to attain comparable SE, thereby reducing computational complexity. These proposed techniques can employ coherent and non-coherent detection and can be adjusted to achieve various spectral efficiencies by altering the number of multiplexed layers. Unlike our proposed LCSS and LDMCSS, other CSS alternatives for LPWANs cannot
provide the same level of flexibility and SE. The performance of these techniques is evaluated in terms of bit error rate under different channel conditions, as well as with phase and frequency offsets.
\end{abstract}
\IEEEpeerreviewmaketitle

\section{Introduction}
\IEEEPARstart{L}{ow}-power wide-area networks (LPWANs) are crucial for Internet-of-Things (IoT), as they enable communication between battery-powered devices while consuming minimal power. One such LPWAN technology is Long Range (LoRa)-WAN, which uses a proprietary waveform, LoRa as a physical layer modulation. LoRa is a simple form of chirp spread spectrum (CSS) and can balance sensitivity and data rates for fixed channel bandwidths by using different spreading factors (SF) as described in \cite{lorawan_spec,lora_mod_basics}. Semtech corporation, which holds the LoRa patent, has never published the details of the modulation's signal structure design. However, Vangelista in \cite{lora}, M. Chiani, and A. Elzanaty in \cite{chiani2019lora} have explained the waveform design, spectral properties, detection principles, etc. LoRa symbols are defined using \(M\) cyclic time shifts of the chirp, which are used as the information-bearing elements. These cyclic time shifts correspond to frequency shifts (FSs) of the complex conjugate of the chirp signal, i.e., down-chirp signal; thus, LoRa can be considered FS chirp modulation. The scalable parameter that allows for different spectral efficiencies and energy efficiencies for LoRa is the SF, i.e., \(\lambda = \log_2(M)\), where \(\lambda = \llbracket 7,12\rrbracket \). \(\lambda\) is equivalent to the number of bits a LoRa symbol can transmit in a symbol duration of \(T_\mathrm{s}\).

LoRa is widely adopted but has a relatively low achievable spectral efficiency (SE) across the bands it operates in. As a result, alternatives that are more spectrally efficient such as CSS schemes like \cite{ics_lora,e_lora,psk_lora, do_css,ssk_lora,iqcss,dcrk_css,fscss_im,iqcim,ssk_ics_lora,epsk_lora,gcss,tdm_lora,dm_css,azim2022dual} have been explored. The waveform design of these CSS variants has been studied in-depth in \cite{azim2022survey}, which include some schemes that use FSs in both the in-phase and quadrature components, like \cite{e_lora,iqcss,iqcim}. In contrast, others employ phase shifts (PSs) like in \cite{psk_lora, epsk_lora}. Some also use the index modulation, for example, \cite{fscss_im,iqcim}. While these variants generally have better spectral efficiencies than LoRa, they also have limitations.

The most spectral-efficient schemes among those listed are the in-phase and quadrature time-domain multiplexed (IQ-TDM)-CSS \cite{tdm_lora} and the dual-mode (DM)-TDM-CSS \cite{azim2022dual}. IQ-TDM-CSS offers four times higher SE relative to the classical LoRa while being more energy-efficient than it. However, due to the use of both the in-phase and the quadrature domains, non-coherent detection is not possible. Moreover, IQ-TDM-CSS is also very sensitive to phase and frequency offsets making it less practical. In contrast, DM-TDM-CSS rectifies the limitation of IQ-TDM-CSS by permitting forthright non-coherent, coherent detection processes while being resilient to the phase and frequency offsets. Nonetheless, for both approaches, the alphabet cardinality, \(M\) is the only tunable parameter to attain different spectral or energy efficiencies. To be more precise, IQ-TDM-CSS and DM-TDM-CSS can transmit \(4\lambda\) and \(4\lambda-4\) bits per \(T_\mathrm{s}\). Thus, for \(\lambda=\llbracket 7,12 \rrbracket\), a limited number of points are available on the SE versus energy (EE) performance curve. This indicates the achievability of a limited number of spectral efficiencies, where the maximum achievable SE for IQ-TDM-CSS and DM-TDM-CSS is \(\sfrac{4\lambda}{M}\) bits/s/Hz and \(\sfrac{4\lambda-4}{M}\) bits/s/Hz, respectively. Some approaches like slope-shift keying interleaved chirp spreading (SSK-ICS)-LoRa, DM-CSS, and discrete chirp rate keying (DCRK)-CSS are also more efficient than classical LoRa; however, their maximum achievable SE is considerably less than IQ-TDM-CSS and DM-TDM-CSS. Since this work aims to propose approaches capable of achieving high SE that are also adaptable in terms of achieving different spectral and energy efficiencies; therefore, we consider IQ-TDM-CSS and DM-TDM-CSS for comparison. We also consider TDM-CSS \cite{tdm_lora} for comparison because, like the proposed approaches, it also classifies as TDM modulation. The SE of TDM-CSS is \(\sfrac{2\lambda}{M}\) bits/s/Hz, which is twice that of LoRa.

The article proposes layered CSS (LCSS) and layered DMCSS (LDMCSS), that enhance the performance of TDM-CSS and DM-TDM-CSS, respectively, by utilizing time-domain multiplexing to combine symbols with different chirp rates. LCSS multiplexes symbols with varying chirp rates from LoRa, while LDMCSS multiplexes symbols with varying chirp rates from a modified version of DM-CSS that does not incorporate the PSs. Unlike LDMCSS, which uses one even and one odd FS for the un-chirped symbol in each layer, LCSS employs only one activated FS per un-chirped symbol in each layer. This results in LDMCSS requiring half the number of layers compared to LCSS to achieve similar SE, reducing the transceiver complexity by half. The proposed waveform design combines the time-domain multiplexing design with the existing schemes to develop a layered signal structure. But at the same time,  the proposed approaches demonstrate remarkable adaptability to obtain varying spectral and energy efficiency, allowing for customization according to specific application requirements. They differ from existing waveform designs due to their degree of flexibility. It is highlighted that the the layered CSS schemes offer an extra degree of freedom that LoRa does not have, which is beneficial when a more diverse set of SE is desired. Therefore, these schemes have a lot of potential as workable alternatives to established waveform designs for LPWANs. 

In the subsequent sections of this article, we will comprehensively depict the waveform design of LCSS and LDMCSS schemes. Following the study of the transmission characteristics, we will furnish the coherent and non-coherent detection mechanisms. Furthermore, we will analyze the interference that arises due to the layered signal structure of the proposed schemes. The layered signal structure of the proposed schemes engenders inter-layer interference that can diminish the performance of the schemes. Additionally, we will analyze the orthogonality of the proposed schemes. 
We will also present simulation and numerical results to demonstrate the performance of the proposed schemes. The results demonstrate that the proposed schemes surpass the classical approaches while providing substantially higher SE and improved EE. Additionally, we will illustrate that the proposed schemes can be adjusted to provide various spectral and energy efficiencies by adding or removing layers in the signal structure at the cost of increased or reduced computational complexity.

Against the given background, the contributions of this work are as follows:
\begin{enumerate}
\item We  \textit{propose LCSS and LDMCSS schemes} as alternatives to state-of-the-art CSS schemes (including LoRa) for LPWANs. We demonstrate that the layered signal structures result in energy and spectral-efficient schemes, that are robust against frequency and phase offsets. 
\item We thoroughly \textit{elucidate the transceiver architecture} of LCSS and LDMCSS by providing a detailed description of the waveform generation process and the coherent and non-coherent detection mechanisms. This includes a clear explanation of the steps involved in the generation and detection processes and the design choices made for each scheme.
\item We \textit{analytically examine the interference} that transpires due to the superposition of time-domain multiplexed symbols in LCSS and LDMCSS. The analysis indicates that interference increases with the number of layers. The orthogonality analysis, on the other hand, illustrates that the symbols lose their orthogonality due to the layered structure. These analysis proffers an understanding of the manner in which the interference impacts the quality of the signal and the efficacy of the detection mechanisms.
\item We provide \textit{theoretical BER expressions} for the proposed LCSS and LDMCSS schemes.
\item The article conducts an \textit{assessment and comparison of the performance of the proposed LCSS and LDMCSS with benchmark schemes} by utilizing a set of performance metrics, specifically: (i) SE versus required signal-to-noise ratio (SNR) per bit for a target bit error rate (BER) of \(10^{-3}\);  (ii) BER performance in an additive white Gaussian noise (AWGN) and a frequency-selective fading channel; and (iii) BER performance considering phase and frequency offsets.
\end{enumerate}

The remainder of the article is organized as follows: In Section II presents a comprehensive elucidation of the system model and the transceiver architecture pertaining to the LCSS and LDMCSS. Section III conducts a scrutiny of the orthogonality of the schemes proposed. Subsequently, in Section IV, an examination of the interference arising from the layered signal configuration within the proposed schemes. The theoretical BER expressions for the proposed schemes are provided in Section V for non-coherent detection and upper-bound BER expressions are developed for coherent detection. The outcomes of simulated results  are presented in Section VI, and finally, in Section VII, conclusions are drawn based upon the results gleaned in Section VI.
\section{Proposed Layered CSS Modulation Schemes}
In this section, we first provide a system model. After that, we present the transmission and detection mechanisms of  LCSS and LDMCSS. To design the transceiver architecture, our approach simply extends the well-known methodologies of the receiver design to the new transmission scheme in order to keep the complexity low, given that the maximum likelihood implementation of the layered CSS schemes would have considerably higher complexity.
\subsection{System Model}
We consider a chirped CSS symbol is composed of two primary elements: (i) an un-chirped symbol, \(g(n)\), where \(n=\llbracket 0, M-1\rrbracket\); and (ii) a spreading symbol, \(c(n)\) that spreads the information in the bandwidth, \(B=\sfrac{M}{T_\mathrm{s}}\), where \(T_\mathrm{s}\) is the symbol period. The un-chirped symbol can have one (as in LoRa), or multiple (as in DM-CSS) activated FSs. Spreading results in an injective mapping of FSs to cyclic time shifts. It is noteworthy that spreading on an un-chirped symbol can have different slope/chirp rates \cite{ssk_lora, dcrk_css}. Here, by different chirp rates, we mean the normalized chirp rate over the bandwidth. Specifically, LoRa chirp rate is defined as \(a = \sfrac{B}{T_\mathrm{s}}\) Hz/s, such that in \(T_\mathrm{s}\) seconds the signal spans \(B\) Hz. In our case, we generalize the chirp rate to be \(a = \sfrac{lB}{T_\mathrm{s}}\). Notice that this does not imply that our signal spans \(l B\) Hz in \(T_\mathrm{s}\) seconds because the signal is wrapped around the limits \(\left[-\sfrac{B}{2},\sfrac{B}{2}\right]\). Basically, we can increase the number of bits/s/Hz if we have multiple signals with different chirp rates in the same bandwidth, which is not possible in the typical single-user LoRa transmission. 

\sloppy{In the discrete time, the CSS symbol consists of \(M\) samples and is given as \(s(n)=g(n)c(n)\). Then, the discrete-time baseband received symbol assuming perfect time and frequency synchronization for simplicity\footnote{The numerical evaluation of these schemes consider phase and frequency offsets.} is given as:}
\begin{equation}\label{eq1}
y(n) = hs(n) + w(n),
\end{equation}
where \(h\) is the complex channel gain, and \(w(n)\) corresponds to AWGN samples. We are considering AWGN and its characterization in terms of its single-sided noise power spectral density, \(N_0\), and noise variance, \(\sigma_n^2=N_0B\). In LPWANs, CSS symbols maintain a narrow bandwidth of \(500\) kHz or smaller; therefore, a flat fading channel can have constant attenuation over the entire \(B\). 
\subsection{Layered Chirp Spread Spectrum}
\begin{figure}[t]\centering
\includegraphics[trim={0 0 0 0},clip,scale=0.9]{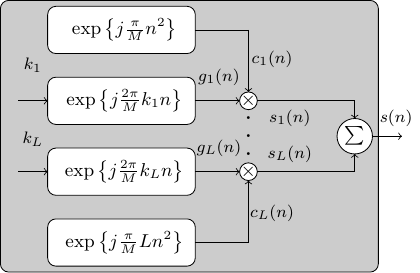}
  \caption{LCSS transmitter architecture. }
\label{fig1}
\end{figure}
\begin{figure}[t]\centering
\includegraphics[trim={0 0 0 0},clip,scale=0.9]{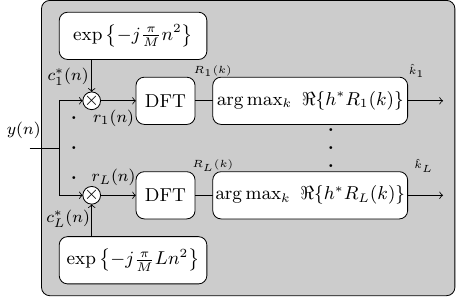}
  \caption{LCSS coherent receiver architecture. }
\label{fig2}
\end{figure}


\subsubsection{Transmission}
Fig. \ref{fig1} shows the generalized transmitter architecture of LCSS. LCSS uses \(M_\mathrm{c}\) (positive) chirp rates, and the number of layers is \(L = M_\mathrm{c}\) (which is assumed to be a multiple of two for simplicity). The un-chirped symbol of the \(l\)th layer with one activated FS is \(g_l(n)= \exp\left\{j\frac{\pi}{M}2k_l n\right\}\), where \(l= \llbracket 1, L \rrbracket\), \(k_l \in \left[ 0, M-1\right]\) and \(n= \llbracket 0, M-1\rrbracket\). Note that the \(k_l\) is attained after the binary-to-decimal conversion of \(\lambda_l=\log_2(M)\) bits. The spreading symbol of the \(l\)th layer, \(c_l(n)= \exp\left\{j\frac{\pi}{M}l n^2\right\}\) is then used to attain the \(l\)th layer chirped symbol, i.e.,:
\begin{equation}\label{eq2}
s_l(n)= g_l(n)c_l(n)=\exp\left\{j\frac{\pi}{M}\left(2k_ln+ln^2\right)\right\}.
\end{equation}

Subsequently, the \(L\) chirped symbols are added to attain a composite signal \(s(n)\) as:
\begin{equation}\label{eq3}
s(n)= \sum_{l=1}^{L} s_l(n)=\sum_{l=1}^{L} \exp\left\{j\frac{\pi}{M}\left(2k_ln+ln^2\right)\right\}.
\end{equation}

The symbol energy of \(s(n)\) is equal to \(E_\mathrm{s}=\mathbb{E}\{\vert s(n)\vert^2\}=\sum_{n=0}^{M-1}\mathbb{E}\{\vert s(n)\vert^2\}\), and the number of bits it transmits is \(\overline{\lambda} = \sum_{l=1}^{L}\lambda_l=L\lambda\). Here, \(\mathbb{E}\{\cdot\}\) denotes the expectation operator.
\subsubsection{Detection}
For clarity of exposition, we consider the following vectorial representations,  \(\boldsymbol{y}= \left[y(0),y(1),\cdots,y(M-1)\right]^\mathrm{T}\), and \(\boldsymbol{s}= \left[s(0),s(1),\cdots,s(M-1)\right]^\mathrm{T}\), where \([\cdot]^\mathrm{T}\) is the transpose operator. 
%
%

The coherent detector for LCSS is depicted in Fig. \ref{fig2}. Considering that channel state information (CSI), \(h\) is available at the receiver and the transmit symbols are equiprobable, the coherent detection dictates maximizing the probability of receiving \(\boldsymbol{y}\) when \(\boldsymbol{s}\) was sent given \(h\), i.e., \(\mathrm{prob}\left(\boldsymbol{y}\vert \boldsymbol{s},h\right)\). 
Then, the coherent detection problem given as:
\begin{equation}\label{eq4}
\begin{split}
\left\{\hat{k}_l\right\}_{l=1}^{L} =\mathrm{arg}\max_{k_l} ~\mathrm{prob}\left(\boldsymbol{y}\vert \boldsymbol{s},h\right) =\mathrm{arg}\max_{k_l}~\Re\left\{\langle \boldsymbol{y},h\boldsymbol{s}\rangle\right\}.
\end{split}
\end{equation}

Considering that \((\cdot)^\ast\) evaluates the complex conjugate, \(\langle \boldsymbol{y},h\boldsymbol{s}\rangle\) is  simplified as:
\begin{equation}\label{eq5}
\begin{split}
\langle \boldsymbol{y},h\boldsymbol{s}\rangle & ={h}^\ast \sum_{n=0}^{M-1}\sum_{l=1}^{L}y(n){s}_l^\ast(n)= {h}^\ast \sum_{n=0}^{M-1}\sum_{l=1}^{L}y(n)g_l^\ast(n)c_l^\ast(n)\\&= {h}^\ast \sum_{n=0}^{M-1}\sum_{l=1}^{L} r_l(n)g_l^\ast(n)=  {h}^\ast \sum_{l=1}^{L} R_l(k),
\end{split}
\end{equation}
for \(k = \llbracket 0, M-1\rrbracket\), where \(r_l(n) = y(n)c_l^\ast(n)\). \(R_l(k)\) is the discrete Fourier transform (DFT) of \(r_l(n)\). The simplification of \(\langle \boldsymbol{y},h\boldsymbol{s}\rangle\) in (\ref{eq5}), leads to:
\begin{equation}\label{eq6}
\begin{split}
\left\{\hat{k}_l\right\}_{l=1}^{L} =\mathrm{arg}\max_{k_l} ~\Re\left\{{h}^\ast \sum_{l=1}^{L} R_l(k)\right\}.
\end{split}
\end{equation}

The FSs can also be dis-jointly identified as:
\begin{equation}\label{eq7}
\begin{split}
\hat{k}_l =\mathrm{arg}\max_{k_l} ~\Re\left\{{h}^\ast R_l(k)\right\}.
\end{split}
\end{equation}

For coherent detection, the DFT of the un-chirped symbol is first computed using the chirp rate respective to each layer, i.e., \(R_l(k)\). Then, the FS of the same layer is identified by evaluating the maximum argument of \(\Re\{R_l(k)\}\) multiplied by the conjugate of \(h\).

\begin{figure}[t]\centering
\includegraphics[trim={0 0 0 0},clip,scale=1]{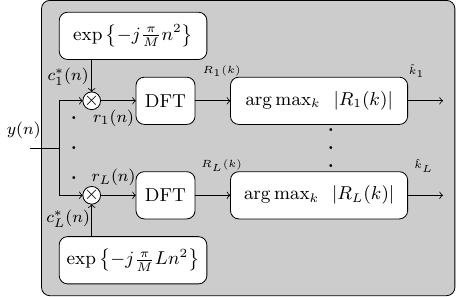}
  \caption{LCSS non-coherent receiver architecture. }
\label{fig3}
\end{figure}
%
%

When CSI, \(h\) is unavailable, the non-coherent detection mechanism is generally employed. The computational complexity of the non-coherent detector is notably less than that of coherent detection, resulting in low-power consumption and low-cost LPWAN components. The non-coherent detector for LCSS is illustrated in Fig. \ref{fig3}. For non-coherent detection, the FSs of the \(l\)th layer are identified dis-jointly as: 
\begin{equation}\label{eq8}
\begin{split}
\hat{k}_l =\mathrm{arg}\max_{k_l} ~\left\vert R_l(k)\right\vert.
\end{split}
\end{equation}

Note that (\ref{eq8}) follows (\ref{eq5}) as the non-coherent detector evaluates the maximum argument of \(\left \vert \sum_{n=0}^{M-1}y(n){s}^\ast(n) \right\vert\). The DFT of \(r_l(n)\) is evaluated, which yields \(R_l(k)\). Then, the FS of the \(l\)th layer is identified by determining the maximum argument of \(\vert R_l(k) \vert \) as in (\ref{eq8}).

%
\subsection{Layered Dual-Mode Chirp Spread Spectrum}

\begin{figure}[t]\centering
\includegraphics[trim={0 0 0 0},clip,scale=0.93]{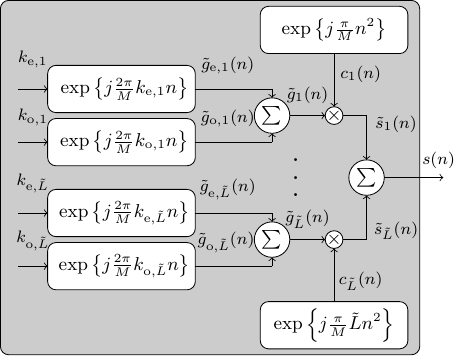}
  \caption{LDMCSS transmitter architecture. }
\label{fig4}
\end{figure}
\subsubsection{Transmission}
%
Unlike LCSS, where only one FS was activated, in LDMCSS, each layer activates two FSs, one even and one odd, as in the DM-CSS, thus halving the number of layers relative to LCSS to attain almost the same SE. Let \(\tilde{L}\) be the number of layers for LDMCSS; then, we have \(\tilde{L} = \sfrac{L}{2}\).  Fig. \ref{fig4} illustrates the transmitter architecture of LDMCSS. For the un-chirped symbol at each layer, a total of \(M\) FSs are available; half are even, and half are odd. For the \(\tilde{l}\)th layer, where \(\tilde{l}= \llbracket 1, \tilde{L}\rrbracket\), the even and the odd activated FSs, \(k_{\mathrm{e},\tilde{l}}\) and \(k_{\mathrm{o},\tilde{l}}\), are attained after binary-to-decimal conversion of \(\tilde{\lambda}_{\mathrm{e},\tilde{l}}= \log_2(\sfrac{M}{2})\) and \(\tilde{\lambda}_{\mathrm{o},\tilde{l}}= \log_2(\sfrac{M}{2})\) bits, where \(k_{\mathrm{e},\tilde{l}}\in \left[0,\sfrac{M}{2}-1\right]\) and \(k_{\mathrm{o},\tilde{l}}\in \left[0,\sfrac{M}{2}-1\right]\). Since the number of layers in LDMCSS are half relative to LCSS, the number of chirp rates needed is also half in LDMCSS compared to LCSS, i.e., \(\tilde{M}_\mathrm{c}= \sfrac{M_\mathrm{c}}{2}\). The un-chirped symbol for the \(\tilde{l}\)th layer in LDMCSS is  \(\tilde{g}_{\tilde{l}}(n)= \tilde{g}_{\mathrm{e},\tilde{l}}(n) + \tilde{g}_{\mathrm{o},\tilde{l}}(n) =\exp\left\{j\frac{\pi}{M}2k_{\mathrm{e},\tilde{l}}~ n\right\}+\exp\left\{j\frac{\pi}{M}2k_{\mathrm{o},\tilde{l}}~ n\right\}\). Then using the spreading symbol for the \(\tilde{l}th\) layer,  \({c}_{\tilde{l}}(n)= \exp\left\{j\frac{\pi}{M}\tilde{l}~ n^2\right\}\), the chirped symbol for the \(\tilde{l}\)th layer is as follows:
\begin{equation}\label{eq9}
\begin{split}
\tilde{s}_l(n) = \tilde{g}_{\tilde{l}}(n)c_{\tilde{l}}(n) &= \exp\left\{j\frac{\pi}{M}\left(2k_{\mathrm{e},\tilde{l}}~n+\tilde{l}n^2\right)\right\}\\&~+\exp\left\{j\frac{\pi}{M}\left(2k_{\mathrm{o},\tilde{l}}~n+\tilde{l}n^2\right)\right\}.
\end{split}
\end{equation}

The composite chirped symbol of \(\tilde{L}\) layers is attained as:
\begin{equation}\label{eq10}
\begin{split}
s(n) =\sum_{\tilde{l}=0}^{\tilde{L}}\tilde{s}_{\tilde{l}}(n) &= \sum_{\tilde{l}=0}^{\tilde{L}}\exp\left\{j\frac{\pi}{M}\left(2k_{\mathrm{e},\tilde{l}}~n+\tilde{l}n^2\right)\right\}\\&~+\sum_{\tilde{l}=0}^{\tilde{L}}\exp\left\{j\frac{\pi}{M}\left(2k_{\mathrm{o},\tilde{l}}~n+\tilde{l}n^2\right)\right\}.
\end{split}
\end{equation}

The symbol energy of the LDMCSS transmit symbol, \(s(n)\) is \(E_\mathrm{s}=\mathbb{E}\{\vert s(n)\vert^2\}=\sum_{n=0}^{M-1}\mathbb{E}\{\vert s(n)\vert^2\}\), and the total number of bits transmitted per symbol is \(\tilde{\lambda} = \sum_{\tilde{l}=1}^{\tilde{L}}\left(\tilde{\lambda}_{\mathrm{e},\tilde{l}}+\tilde{\lambda}_{\mathrm{o},\tilde{l}}\right)=\tilde{L}(2\lambda-2)\). Note that the number of bits transmitted per symbol in LDMCSS is marginally less than those in LCSS. For example, if we consider SF of \(\lambda\), then for \(L= 4\) and \(\tilde{L} = 2\), the number of bits transmitted per symbol in LCSS and LDMCSS is \(\overline{\lambda} = 4\lambda\) and \(\tilde{\lambda}=4\lambda-4\), respectively. 
\subsubsection{Detection}
\begin{figure}[t]\centering
\includegraphics[trim={0 0 0 0},clip,scale=0.93]{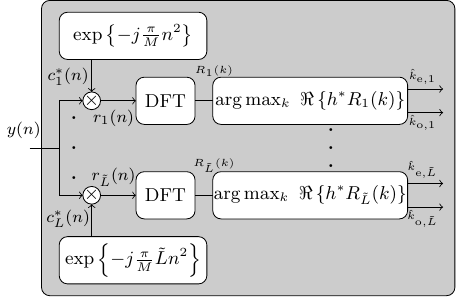}
  \caption{LDMCSS coherent receiver architecture. }
\label{fig5}
\end{figure}

The coherent detector for LDMCSS is illustrated in Fig. \ref{fig5}. The coherent detection problem for LDMCSS is given as:
\begin{equation}\label{eq11}
\begin{split}
\left\{\hat{k}_{\mathrm{e},\tilde{l}},\hat{k}_{\mathrm{o},\tilde{l}}\right\}_{l=1}^{L} &=\mathrm{arg}\max_{k_{\mathrm{e}},k_{\mathrm{o}}} ~\mathrm{prob}\left(\boldsymbol{y}\vert \boldsymbol{s},h\right)=\mathrm{arg}\max_{k_{\mathrm{e}},k_{\mathrm{o}}}~\Re\left\{\langle \boldsymbol{y},h\boldsymbol{s}\rangle\right\},
\end{split}
\end{equation}
where \(k_\mathrm{e}\) and \(k_\mathrm{o}\) are the even and the odd FSs. \(\langle \boldsymbol{y},h\boldsymbol{s}\rangle\) in (\ref{eq11}) simplifies to: 
\begin{equation}\label{eq12}
\begin{split}
\langle \boldsymbol{y},h\boldsymbol{s}\rangle & = {h}^\ast \sum_{n=0}^{M-1}\sum_{\tilde{l}=1}^{\tilde{L}}y(n)\tilde{s}_{\tilde{l}}^\ast(n)\\&= {h}^\ast \left(\sum_{n=0}^{M-1}\sum_{\tilde{l}=1}^{\tilde{L}}r_{\tilde{l}}(n)g_{\mathrm{e},\tilde{l}}^\ast(n)+\sum_{n=0}^{M-1}\sum_{\tilde{l}=1}^{\tilde{L}}r_{\tilde{l}}(n)g_{\mathrm{o},\tilde{l}}^\ast(n)\right)\\&=  {h}^\ast \sum_{\tilde{l}=1}^{\tilde{L}} \biggl(R_{\tilde{l}}(k_{\mathrm{e}})+ R_{\tilde{l}}(k_{\mathrm{o}})\biggr).
\end{split}
\end{equation}

Using (\ref{eq12}), the coherent detection problem in (\ref{eq11}) becomes:
\begin{equation}\label{eq13}
\begin{split}
\left\{\hat{k}_{\mathrm{e},\tilde{l}},\hat{k}_{\mathrm{o},\tilde{l}}\right\}_{l=1}^{L} &=\mathrm{arg}\max_{k_{\mathrm{e}},k_{\mathrm{o}}}~\Re\left\{{h}^\ast \sum_{\tilde{l}=1}^{\tilde{L}} \biggl(R_{\tilde{l}}(k_{\mathrm{e}})+ R_{\tilde{l}}(k_{\mathrm{o}})\biggr)\right\}.
\end{split}
\end{equation}

The even and the odd FSs of the \(\tilde{l}\)th layer are attained in a dis-joint manner as:
\begin{equation}\label{eq14}
\begin{split}
\hat{k}_{\mathrm{e},\tilde{l}} =\mathrm{arg}\max_{k_{\mathrm{e}}}~\Re\left\{{h}^\ast R_{\tilde{l}}(k)\right\},
\end{split}
\end{equation}
and
\begin{equation}\label{eq15}
\begin{split}
\hat{k}_{\mathrm{o},\tilde{l}} =\mathrm{arg}\max_{k_{\mathrm{o}}}~\Re\left\{{h}^\ast R_{\tilde{l}}(k)\right\},
\end{split}
\end{equation}
%
%
respectively. We attain the even or odd FS by determining the real component of the DFT of \(r_{\tilde{l}}\), i.e., \(R_{\tilde{l}}\) multiplied by \(h^\ast\). 
%
%
\begin{figure}[t]\centering
\includegraphics[trim={0 0 0 0},clip,scale=0.93]{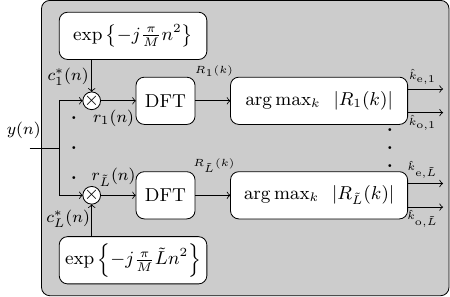}
  \caption{LDMCSS non-coherent receiver architecture. }
\label{fig6}
\end{figure}
We obtain the non-coherent detector for LDMCSS by evaluating the maximum argument of \(\left\vert \sum_{n=0}^{M-1}y(n){s}^\ast(n)\right\vert\), leading to the following dis-joint detection problem for the identification of even and odd FSs:
\begin{equation}\label{eq16}
\begin{split}
\hat{k}_{\mathrm{e},\tilde{l}} =\mathrm{arg}\max_{k_{\mathrm{e}}}~\left\vert R_{\tilde{l}}(k)\right\vert,
\end{split}
\end{equation}
and 
\begin{equation}\label{eq17}
\begin{split}
\hat{k}_{\mathrm{o},\tilde{l}} &=\mathrm{arg}\max_{k_{\mathrm{o}}}~\left\vert R_{\tilde{l}}(k)\right\vert,
\end{split}
\end{equation}
respectively. LDMCSS non-coherent detector is illustrated in Fig. \ref{fig6}. To determine the even or the odd FS using the non-coherent detector, the maximum argument of the absolute of the DFT of \(r_{\tilde{l}}\), i.e., \(R_{\tilde{l}}\). 

%
%
\section{Orthogonality Analysis of the Proposed Schemes}
In this section, we investigate the orthogonality of LCSS and LDMCSS symbols. To this end, we evaluate the inner product of \(\boldsymbol{s}\) and \(\overline{\boldsymbol{s}}=\left[\overline{s}(0),\overline{s}(1),\cdots, \overline{s}(M-1)\right]^\mathrm{T}\) (the same symbol notations are considered for both LCSS and LDMCSS). The inner product is given as:
\begin{equation}\label{eq18}
\langle\boldsymbol{s},\overline{\boldsymbol{s}}\rangle= \sum_{n=0}^{M-1} s(n) \overline{s}^\ast(n).
\end{equation}
\subsection{Layered Chirp Spread Spectrum}
For LCSS, we consider that the activated FS of a given layer in \(\boldsymbol{s}\) and \(\overline{\boldsymbol{s}}\) are \(k_l\) and \(\overline{k}_{\overline{l}}\), respectively. \(\boldsymbol{s}\) and \(\overline{\boldsymbol{s}}\) have \(L\) and \(\overline{L}=L\) layers, respectively, with positive chirp rates of \(l\) and \(\overline{l}\), where \(l= \llbracket 1, L\rrbracket\) and \(\overline{l}= \llbracket 1, \overline{L}\rrbracket\). To verify the orthogonality, the two symbols \(\boldsymbol{s}\) and \(\overline{\boldsymbol{s}}\) must have different activated FS, then, the inner product of \(\boldsymbol{s}\) and \(\overline{\boldsymbol{s}}\) yields:
\begin{equation}\label{eq19}
\begin{split}
&\langle\boldsymbol{s},\overline{\boldsymbol{s}}\rangle= \sum_{l=1}^{L}\sum_{\overline{l}=1}^{\overline{L}}\beta_{\left(l,\overline{l}\right)},
\end{split}
\end{equation}
\sloppy{for \(l\neq \overline{l}\), with \(\beta_{\left(l,\overline{l}\right)}= \sum_{n=0}^{M-1}\exp\left\{j\frac{\pi}{M}\biggl(2\kappa_{\left(l,\overline{l}\right)}n+\alpha_{\left(l,\overline{l}\right)}n^2\biggr)\right\}\), where \(\kappa_{\left(l,\overline{l}\right)} = k_l -\overline{k}_{\overline{l}}\), and \(\alpha_{\left(l,\overline{l}\right)} = l - \overline{l}\). Based on the layers, \(l\) and \(\overline{l}\), \(\alpha_{\left(l,\overline{l}\right)}\) ranges between \(-M_\mathrm{c}+1 \leq  \alpha_{\left(l,\overline{l}\right)} \leq M_\mathrm{c}+1\). Since \(\beta_{\left(l,\overline{l}\right)}\) in takes the form of \textit{generalized quadratic Gauss sum}, therefore, we have \(
\beta_{\left(l,\overline{l}\right)}=\theta_{\left(l,\overline{l}\right)}\exp\left\{-j\frac{2\pi}{M}\kappa^2_{\left(l,\overline{l}\right)}\right\}\left(1+\gamma_{\left(l,\overline{l}\right)}\right)\), 
where \(\theta_{\left(l,\overline{l}\right)}=\sqrt{\left\vert \frac{M}{\alpha_{\left(l,\overline{l}\right)}}\right\vert}\exp\left\{j\frac{\pi\vert \alpha_{\left(l,\overline{l}\right)}\vert }{4\alpha_{\left(l,\overline{l}\right)}}\right\}\). Moreover,}
\begin{equation}\label{eq22a}
\begin{split}
\gamma_{\left(l,\overline{l}\right)}&=\sum_{n=0}^{\vert \alpha_{\left(l,\overline{l}\right)}\vert -1}\exp\left\{-j\frac{\pi}{\alpha_{\left(l,\overline{l}\right)}}\left(2\kappa_{\left(l,\overline{l}\right)} n +Mn^2\right)\right\}\\&=
  \begin{dcases}
    1 & \quad \mathrm{rem}\left(\vert \kappa_{\left(l,\overline{l}\right)}\vert,2\right)=0\\
   -1 & \quad \mathrm{rem}\left(\vert \kappa_{\left(l,\overline{l}\right)}\vert,2\right)=1 \\
  \end{dcases}.
\end{split}
\end{equation}
%

Depending on \(k_l\) and \(\overline{k}_{\overline{l}}\), the value of \(\gamma_{\left(l,\overline{l}\right)}\) is either \(1\) or \(-1\).
%
%
where \(\mathrm{rem}\left(x,y\right)\) disseminates the remainder resulting from the division of \(x\) from \(y\). In other words, if \(\vert \kappa_{\left(l,\overline{l}\right)}\vert\) is even, the value of \(\gamma_{\left(l,\overline{l}\right)}\) is \(1\); otherwise, it is \(-1\). Substituting \(\beta_{\left(l,\overline{l}\right)}\) in (\ref{eq19}), we attain:
\begin{equation}\label{eq25}
\begin{split}
\langle\boldsymbol{s},\overline{\boldsymbol{s}}\rangle&= \sum_{l=1}^{L}\sum_{\overline{l}=1}^{\overline{L}}\theta_{\left(l,\overline{l}\right)}\exp\left\{-j\frac{2\pi}{M}\kappa^2_{\left(l,\overline{l}\right)}\right\}\left(1+\gamma_{\left(l,\overline{l}\right)}\right),
\end{split}
\end{equation}
\sloppy{for \(l\neq \overline{l}\). From (\ref{eq25}), we observe that the LCSS symbols are not orthogonal because of the activation of different FSs in the layered structure. Even though the individual layers (with different activated FSs) may be orthogonal to each other; however, the composite symbol as a whole is not orthogonal to another LCSS symbol with different activated FSs. The two symbols can only be orthogonal if and only if all \(\vert\kappa_{\left(l,\overline{l}\right)}\vert\) is odd \(\forall~\{l,\overline{l}\}\), as it would result in \(\gamma_{\left(l,\overline{l}\right)}=-1\). }
\subsection{Layered Dual-Mode Chirp Spread Spectrum}
In LDMCSS,  the even and the odd FSs of \(\boldsymbol{s}\) are \(k_{\mathrm{e},\tilde{l}}\) and \(k_{\mathrm{o},\tilde{l}}\), respectively. Similarly, for \(\overline{\boldsymbol{s}}\), the even and the odd FSs are \(\overline{k}_{\mathrm{e},\overline{\tilde{l}}}\) and \(\overline{k}_{\mathrm{o},\overline{\tilde{l}}}\), respectively. We consider that both  \(\boldsymbol{s}\) and \(\overline{\boldsymbol{s}}\) have \(\tilde{L}\) and \(\overline{\tilde{L}}\) layers, respectively, having positive chirp rates of \(\tilde{l}\) and \(\overline{\tilde{l}}\), where \(\tilde{l}= \llbracket 1, \tilde{L}\rrbracket\) and \(\overline{\tilde{l}}= \llbracket 1, \overline{\tilde{L}}\rrbracket\). The following conditions must hold to determine the orthogonality of LDMCSS symbols: (i) the  even FS in the \(l\)th layer of  \(\boldsymbol{s}\), i.e., \(k_{\mathrm{e},\tilde{l}}\) is different from the even FS of \(\overline{l}\)th layer of \(\overline{\boldsymbol{s}}\), i.e., \(\overline{k}_{\mathrm{e},\overline{\tilde{l}}}\), where \(\tilde{l} = \overline{\tilde{l}}\); and (ii) considering the same layer of \(\boldsymbol{s}\) and \(\overline{\boldsymbol{s}}\), i.e., \(\tilde{l} = \overline{\tilde{l}}\), the odd FS in \(\boldsymbol{s}\), \(k_{\mathrm{o},\tilde{l}}\) should be different from the odd FS of \(\overline{\boldsymbol{s}}\), \(\overline{k}_{\mathrm{o},\overline{\tilde{l}}}\). The inner product of \(\boldsymbol{s}\) and \(\overline{\boldsymbol{s}}\) for LDMCSS is given as:
\begin{equation}\label{eq26}
\begin{split}
\langle\boldsymbol{s},\overline{\boldsymbol{s}}\rangle= \sum_{l=1}^{L}\sum_{\overline{l}=1}^{\overline{L}}\biggl({\beta}_{\mathrm{e}\left(\tilde{l},\overline{\tilde{l}}\right)}+{\beta}_{\mathrm{o}\left(\tilde{l},\overline{\tilde{l}}\right)}\biggr),
\end{split}
\end{equation}
\sloppy{for \(l\neq \overline{l}\), where $\small{
{\beta}_{\mathrm{e}\left(\tilde{l},\overline{\tilde{l}}\right)}= \sum_{n=0}^{M-1}\exp\left\{j\frac{\pi}{M}\biggl(2\kappa_{\mathrm{e}\left(\tilde{l},\overline{\tilde{l}}\right)}n+\tilde{\alpha}_{\left(\tilde{l},\overline{\tilde{l}}\right)}n^2\biggr)\right\}}$, and \(\small{{\beta}_{\mathrm{o}\left(\tilde{l},\overline{\tilde{l}}\right)}= \sum_{n=0}^{M-1}\exp\left\{j\frac{\pi}{M}\biggl(2\kappa_{\mathrm{o}\left(\tilde{l},\overline{\tilde{l}}\right)}n+\tilde{\alpha}_{\left(\tilde{l},\overline{\tilde{l}}\right)}n^2\biggr)\right\}}\), where \(\kappa_{\mathrm{e}\left(\tilde{l},\overline{\tilde{l}}\right)}=k_{\mathrm{e},\tilde{l}}-\overline{k}_{\mathrm{e},\overline{\tilde{l}}}\), \(\kappa_{\mathrm{o}\left(\tilde{l},\overline{\tilde{l}}\right)}=k_{\mathrm{o},\tilde{l}}-\overline{k}_{\mathrm{o},\overline{\tilde{l}}}\), and \(\tilde{\alpha}_{\left(\tilde{l},\overline{\tilde{l}}\right)}=\tilde{l}-\overline{\tilde{l}}\). Note that \(\tilde{\alpha}_{\left(\tilde{l},\overline{\tilde{l}}\right)}\) is bounded between \(-\tilde{M}_\mathrm{c}+1 \leq \tilde{\alpha}_{\left(\tilde{l},\overline{\tilde{l}}\right)}\leq \tilde{M}_\mathrm{c}-1\). It is recalled that the number of chirp rates in LDMCSS is half relative to the chirp rates used in LCSS, i.e., \(M_\mathrm{c} = \sfrac{\tilde{M}_\mathrm{c}}{2}\).}

We can attain closed-form expressions for \({\beta}_{\mathrm{e}\left(\tilde{l},\overline{\tilde{l}}\right)}\) and \({\beta}_{\mathrm{o}\left(\tilde{l},\overline{\tilde{l}}\right)}\) using the generalized quadratic Gauss sum as 
\(\small{
\beta_{\mathrm{e}\left(\tilde{l},\overline{\tilde{l}}\right)}=\tilde{\theta}_{\left(\tilde{l},\overline{\tilde{l}}\right)}\exp\left\{-j\frac{2\pi}{M}\kappa_{\mathrm{e}^2\left(\tilde{l},\overline{\tilde{l}}\right)}\right\}\left(1+\gamma_{\mathrm{e}\left(l,\overline{l}\right)}\right)}\), and \(\small{\beta_{\mathrm{o}\left(\tilde{l},\overline{\tilde{l}}\right)}=\tilde{\theta}_{\left(\tilde{l},\overline{\tilde{l}}\right)}\exp\left\{-j\frac{2\pi}{M}\kappa_{\mathrm{o}^2\left(\tilde{l},\overline{\tilde{l}}\right)}\right\}\left(1+\gamma_{\mathrm{o}\left(l,\overline{l}\right)}\right)}\), respectively, where \(\tilde{\theta}_{\left(\tilde{l},\overline{\tilde{l}}\right)} = \sqrt{\left\vert \frac{M}{\tilde{\alpha}_{\left(\tilde{l},\overline{\tilde{l}}\right)}}\right\vert}\exp\left\{j\frac{\pi\vert \tilde{\alpha}_{\left(\tilde{l},\overline{\tilde{l}}\right)}\vert }{4\tilde{\alpha}_{\left(\tilde{l},\overline{\tilde{l}}\right)}}\right\}\). Let us consider a parameter \(\gamma_{(\cdot)\left(l,\overline{l}\right)}\), which can be used for either \(\gamma_{\mathrm{e}\left(l,\overline{l}\right)}\) and \(\gamma_{\mathrm{o}\left(l,\overline{l}\right)}\). \(\gamma_{(\cdot)\left(l,\overline{l}\right)}\) is given as: 
\begin{equation}\label{eq32}
\begin{split}
\gamma_{(\cdot)\left(\tilde{l},\overline{\tilde{l}}\right)}&=\sum_{n=0}^{\vert \tilde{\alpha}_{\left(\tilde{l},\overline{\tilde{l}}\right)}\vert -1}\exp\left\{-j\frac{\pi}{\tilde{\alpha}_{\left(\tilde{l},\overline{\tilde{l}}\right)}}\left(2\kappa_{(\cdot)\left(\tilde{l},\overline{\tilde{l}}\right)} n +Mn^2\right)\right\}\\&= \begin{dcases}
    1 & \quad \mathrm{rem}\left(\vert \kappa_{(\cdot)\left(\tilde{l},\overline{\tilde{l}}\right)}\vert,2\right)=0\\
   -1 & \quad \mathrm{rem}\left(\vert \kappa_{(\cdot)\left(\tilde{l},\overline{\tilde{l}}\right)}\vert,2\right)=1 \\
  \end{dcases},
  \end{split}
\end{equation}
%
\(\gamma_{\mathrm{e}\left(\tilde{l},\overline{\tilde{l}}\right)}\) and \(\gamma_{\mathrm{o}\left(\tilde{l},\overline{\tilde{l}}\right)}\) can be either \(1\) or \(-1\) depending on \(\kappa_{\mathrm{e}\left(\tilde{l},\overline{\tilde{l}}\right)}\)  and \(\kappa_{\mathrm{o}\left(\tilde{l},\overline{\tilde{l}}\right)}\). Then, \(\langle\boldsymbol{s},\overline{\boldsymbol{s}}\rangle\) can be obtained as:
\begin{equation}\label{eq36}
\begin{split}
\langle\boldsymbol{s},\overline{\boldsymbol{s}}\rangle &= \sum_{l=1}^{L}\sum_{\overline{l}=1}^{\overline{L}} \tilde{\theta}_{\left(\tilde{l},\overline{\tilde{l}}\right)}\left(\exp\left\{-j\frac{2\pi}{M}\kappa_{\mathrm{e}^2\left(\tilde{l},\overline{\tilde{l}}\right)}\right\}\left(1+\gamma_{\mathrm{e}\left(l,\overline{l}\right)}\right)\right.\\&~~~\left.+\exp\left\{-j\frac{2\pi}{M}\kappa_{\mathrm{o}^2\left(\tilde{l},\overline{\tilde{l}}\right)}\right\}\left(1+\gamma_{\mathrm{o}\left(l,\overline{l}\right)}\right)\right),
\end{split}
\end{equation}
for \(\tilde{l}\neq \overline{\tilde{l}}\). From (\ref{eq36}), we gather that, like LCSS, LDMCSS symbols are also not orthogonal even though the individual layers (with different activated even and odd FSs) are orthogonal to each other. This loss in orthogonality is due to the activation of multiple activated even and odd FSs. 
\section{Interference Analysis of the Proposed Schemes}
In this section, we analytically evaluate the interference for LCSS, and LDMCSS caused due to their layered structure. The interference results from activating multiple FSs in the different layers. The analysis considers a simple AWGN channel without loss of generality. To this end, we consider that the received symbol after multiplication with a down-chirp symbol having a chirp rate of \(\overline{l}\) is
\begin{equation}\label{eq37}
\begin{split}
r_{\overline{l}}(n)&= \bigl(s(n)+ w(n)\bigr)c_{\overline{l}}^\ast(n)= s(n)c_{\overline{l}}^\ast(n) + \overline{w}(n),
\end{split}
\end{equation}
where \(\overline{w}(n)= w(n)c_{\overline{l}}^\ast(n)\) are the samples of AWGN multiplied with the down-chirp symbol.

In the sequel, we present the interference analysis of both LCSS and LDMCSS separately. 
\subsection{Layered Chirp Spread Spectrum}
The down-chirped received symbol for LCSS is given as:
\begin{equation}\label{eq38}
\begin{split}
r_{\overline{l}}(n)= \sum_{l=1}^{L}\exp\left\{j\frac{\pi}{M}\biggl(2k_ln +\alpha_{\left(l,\overline{l}\right)}n^2\biggr)\right\}+ \overline{w}(n).
\end{split}
\end{equation}

It is recalled that in the LCSS symbol, the activated FS at the \(l\)th layer is \(k_l\) and is chirped at rate \(l\). When the chirp rate of the down-chirp symbol and the layer number match, i.e., \(\overline{l} = l\), then from (\ref{eq38}), we have:
\begin{equation}\label{eq39}
\begin{split}
r_{\overline{l}}(n)=& \exp\left\{j\frac{2\pi}{M}k_ln\right\}+\sum_{\substack{l=1 \\ l\neq \overline{l}}}^{L}\exp\left\{j\frac{\pi}{M}\biggl(2k_ln +\alpha_{\left(l,\overline{l}\right)}n^2\biggr)\right\}\\&~+ \overline{w}(n).
\end{split}
\end{equation}

The first term in (\ref{eq39}) follows from the fact that \(l = \overline{l}\), whereas the second term incorporates interference caused by the remaining layers. 

Performing \(M\)-order DFT on \(r_{\overline{l}}(n)\) in (\ref{eq39}) results in:
\begin{equation}\label{eq40}
\begin{split}
R_{\overline{l}}(k)&= \sum_{n=0}^{M-1}\exp\left\{j\frac{\pi}{M}\biggl(2k_l-k\biggr)n\right\}+\sum_{\substack{l=1 \\ l\neq \overline{l}}}^{L}\beta_{1\left(l,\overline{l}\right)}+ \overline{W}(k),
\end{split}
\end{equation}
where, \(\beta_{1\left(l,\overline{l}\right)} = \sum_{n=0}^{M-1}\exp\left\{j\frac{\pi}{M}\biggl(2\kappa_{1\left(l\right)}n +\alpha_{\left(l,\overline{l}\right)}n^2\biggr)\right\}\) with \(\kappa_{1\left(l\right)} = k_l-k\). When \(k =  k_l\), then the first term in (\ref{eq40}) equates to \(M\). Moreover, we can attain a closed-form expression for \(\beta_{1\left(l,\overline{l}\right)}\) using generalized quadratic Gauss sum as \(
\beta_{1\left(l,\overline{l}\right)} = \theta_{\left(l,\overline{l}\right)}\exp\left\{j\frac{2\pi}{M}\kappa^2_{1(l)}\right\}\bigl(1+\gamma_{1\left(l,\overline{l}\right)}\bigr)\)
where
\begin{equation}\label{eq43}
\begin{split}
\gamma_{1\left(l,\overline{l}\right)}&=\sum_{n=0}^{\vert \alpha_{\left(l,\overline{l}\right)}\vert -1}\exp\left\{-j\frac{\pi}{\alpha_{\left(l,\overline{l}\right)}}\left(2\kappa_{1\left(l\right)} n +Mn^2\right)\right\}\\&= 
  \begin{dcases}
    1 & \quad \mathrm{rem}\left(\vert \kappa_{1\left(l\right)}\vert,2\right)=0\\
   -1 & \quad \mathrm{rem}\left(\vert \kappa_{1\left(l\right)}\vert,2\right)=1 \\
  \end{dcases},
\end{split}
\end{equation}

\(\gamma_{1\left(l,\overline{l}\right)}\) is either \(1\) or \(-1\), and its value depends on whether \(\vert \kappa_{1\left(l\right)}\vert \) is even or odd.
%

Finally, (\ref{eq40}) can be re-written as:
\begin{equation}\label{eq45}
\begin{split}
R_{\overline{l}}(k)&= \underbrace{M}_{\mathrm{Signal}}+\underbrace{\sum_{\substack{l=1 \\ l\neq \overline{l}}}^{L}\beta_{1\left(l,\overline{l}\right)}}_{\mathrm{Interference}}+ \underbrace{\overline{W}(k)}_{\mathrm{Noise}},
\end{split}
\end{equation}

We can gather from (\ref{eq45}) that apart from the intended signal at a given layer, the FSs on the remaining layers contribute towards the interference. Moreover, (\ref{eq45}) also implies that increasing \(M\) would increase the signal-to-interference ratio for the same number of layers. 
\subsection{Layered Dual-Mode Chirp Spread Spectrum}
The down-chirped LDMCSS received symbol is given as:
\begin{equation}\label{eq46}
\begin{split}
r_{\overline{l}}(n)&= \sum_{\tilde{l}=1}^{L}\exp\left\{j\frac{\pi}{M}\biggl(2k_{\mathrm{e},\tilde{l}}n +\overline{\alpha}_{\left(\tilde{l},\overline{l}\right)}n^2\biggr)\right\}\\&~+\sum_{\tilde{l}=1}^{L}\exp\left\{j\frac{\pi}{M}\biggl(2k_{\mathrm{o},\tilde{l}}n +\overline{\alpha}_{\left(\tilde{l},\overline{l}\right)}n^2\biggr)\right\}+ \overline{w}(n),
\end{split}
\end{equation}
where \(\overline{\alpha}_{\left(\tilde{l},\overline{l}\right)} = \tilde{l}-\overline{l}\). LDMCSS symbols at \(\tilde{l}\)th layer use a chirp rate of \(\tilde{l}\), and the even and odd FSs are \(k_{\mathrm{e},\tilde{l}}\) and \(k_{\mathrm{o},\tilde{l}}\). When the chirp rate of down-chirp symbol matches the chirp rate used in the \(\tilde{l}\)th layer, i.e., \(\tilde{l}=\overline{l}\), then (\ref{eq46}) can be re-written as:
\begin{equation}\label{eq47}
\begin{split}
r_{\overline{l}}(n)&= \exp\left\{j\frac{2\pi}{M}k_{\mathrm{e},\tilde{l}}\right\}+\exp\left\{j\frac{2\pi}{M}k_{\mathrm{o},\tilde{l}}\right\}\\&~+\sum_{\substack{\tilde{l}=1 \\ \tilde{l}\neq \overline{l}}}^{\tilde{L}}\exp\left\{j\frac{\pi}{M}\biggl(2k_{\mathrm{e},\tilde{l}}n +\overline{\alpha}_{\left(\tilde{l},\overline{l}\right)}n^2\biggr)\right\}\\&~+\sum_{\substack{\tilde{l}=1 \\ \tilde{l}\neq \overline{l}}}^{\tilde{L}}\exp\left\{j\frac{\pi}{M}\biggl(2k_{\mathrm{o},\tilde{l}}n +\overline{\alpha}_{\left(\tilde{l},\overline{l}\right)}n^2\biggr)\right\}+ \overline{w}(n).
\end{split}
\end{equation}

The second and third terms in (\ref{eq47}) indicate the interference caused by the layered structure in LDMCSS. Performing \(M\)-order DFT on \(r_{\overline{l}}(n)\) in (\ref{eq47}), we attain:
\begin{equation}\label{eq48}
\begin{split}
R_{\overline{l}}(k)&= \sum_{n=0}^{M-1}\exp\left\{j\frac{\pi}{M}\biggl(2k_{\mathrm{e},\tilde{l}}-k\biggr)n\right\}+ \\&\sum_{n=0}^{M-1}\exp\left\{j\frac{\pi}{M}\biggl(2k_{\mathrm{o},\tilde{l}}-k\biggr)n\right\}\\&~+\sum_{\substack{\tilde{l}=1 \\ \tilde{l}\neq \overline{l}}}^{\tilde{L}}\biggl(\overline{\beta}_{\mathrm{e}\left(\tilde{l},\overline{l}\right)}+\overline{\beta}_{\mathrm{o}\left(\tilde{l},\overline{l}\right)}\biggr)+\overline{W}(k),
\end{split}
\end{equation}
where \(\overline{\beta}_{\mathrm{e}\left(\tilde{l},\overline{l}\right)}= \sum_{n=0}^{M-1}\exp\left\{j\frac{\pi}{M}\biggl(2\overline{\kappa}_{\mathrm{e}\left(\tilde{l}\right)}n +\overline{\alpha}_{\left(\tilde{l},\overline{l}\right)}n^2\biggr)\right\}\), and \(\overline{\beta}_{\mathrm{o}\left(\tilde{l},\overline{l}\right)}= \sum_{n=0}^{M-1}\exp\left\{j\frac{\pi}{M}\biggl(2\overline{\kappa}_{\mathrm{o}\left(\tilde{l}\right)}n +\overline{\alpha}_{\left(\tilde{l},\overline{l}\right)}n^2\biggr)\right\}\).
%
%
\(\overline{\kappa}_{\mathrm{e}\left(\tilde{l}\right)} = k_{\mathrm{e},\tilde{l}}-k\), and \(\overline{\kappa}_{\mathrm{o}\left(\tilde{l}\right)}= k_{\mathrm{o},\tilde{l}}-k\). The closed-form expressions for \(\overline{\beta}_{\mathrm{e}\left(\tilde{l},\overline{l}\right)}\) and \(\overline{\beta}_{\mathrm{o}\left(\tilde{l},\overline{l}\right)}\) can be obtained by using generalized quadratic Gauss sum as \(
\overline{\beta}_{\mathrm{e}\left(\tilde{l},\overline{l}\right)}= \overline{\theta}_{\left(\tilde{l},\overline{l}\right)}\exp\left\{j\frac{2\pi}{M}\kappa^2_{\mathrm{e}(\tilde{l})}\right\}\biggl(1+\overline{\gamma}_{\mathrm{e}\left(\tilde{l},\overline{l}\right)}\biggr)\), and \(\overline{\beta}_{\mathrm{o}\left(\tilde{l},\overline{l}\right)}= \overline{\theta}_{\left(\tilde{l},\overline{l}\right)}\exp\left\{j\frac{2\pi}{M}\kappa^2_{\mathrm{o}(\tilde{l})}\right\}\biggl(1+\overline{\gamma}_{\mathrm{o}\left(\tilde{l},\overline{l}\right)}\biggr)\), 
where \(\overline{\theta}_{\left(\tilde{l},\overline{l}\right)} = \sqrt{\left\vert \frac{M}{\overline{\alpha}_{\left(\tilde{l},\overline{l}\right)}}\right\vert}\exp\left\{j\frac{\pi\vert \overline{\alpha}_{\left(\tilde{l},\overline{l}\right)}\vert }{4\overline{\alpha}_{\left(\tilde{l},\overline{l}\right)}}\right\}\).
Let us define \(\overline{\gamma}_{(\cdot)\left(\tilde{l},\overline{l}\right)}\) to represent \(\overline{\gamma}_{\mathrm{e}\left(\tilde{l},\overline{l}\right)}\) and \(\overline{\gamma}_{\mathrm{o}\left(\tilde{l},\overline{l}\right)}\). \(\overline{\gamma}_{(\cdot)\left(\tilde{l},\overline{l}\right)}\) is then given as:
\begin{equation}\label{eq54}
\begin{split}
\overline{\gamma}_{(\cdot)\left(\tilde{l},\overline{l}\right)} &= \sum_{n=0}^{\vert \overline{\alpha}_{\left(\tilde{l},\overline{l}\right)}\vert -1}\exp\left\{-j\frac{\pi}{\overline{\alpha}_{\left(\tilde{l},\overline{l}\right)}}\left(2\kappa_{(\cdot)\left(\tilde{l}\right)} n +Mn^2\right)\right\} \\&= \begin{dcases}
    1 & \quad \mathrm{rem}\left(\vert \kappa_{(\cdot)\left(\tilde{l}\right)}\vert,2\right)=0\\
   -1 & \quad \mathrm{rem}\left(\vert \kappa_{(\cdot)\left(\tilde{l}\right)}\vert,2\right)=1 \\
  \end{dcases},
\end{split}
\end{equation}
%
 %
\begin{equation}\label{eq56}
\begin{split}
R_{\overline{l}}(k)=\underbrace{M}_{\mathrm{Signal}}+\underbrace{\sum_{\substack{\tilde{l}=1 \\ \tilde{l}\neq \overline{l}}}^{\tilde{L}}\biggl(\overline{\beta}_{\mathrm{e}\left(\tilde{l},\overline{l}\right)}+\overline{\beta}_{\mathrm{o}\left(\tilde{l},\overline{l}\right)}\biggr)}_{\mathrm{Interference}}+\underbrace{\overline{W}(k)}_{\mathrm{Noise}}.
\end{split}
\end{equation}

The conclusions drawn from (\ref{eq56}) are almost the same as those of (\ref{eq45}). Apart from the intended signal, which includes the identification of even and odd FSs on a given layer, the even and the odd FSs of the remaining layers cause interference that may impact the performance significantly. Moreover, like LCSS, increasing \(M\) would improve the signal-to-interference ratio for the same number of layers. 
\section{Theoretical BER Analysis of Layered Schemes}
In this section, we develop the theoretical BER closed-form expressions for LCSS and LDMCSS schemes considering both non-coherent and coherent detection mechanisms. 
\subsection{Layered Chirp Spread Spectrum}
\subsubsection{Non-Coherent Detection}
For \(L=1\) (classical LoRa case), according to \cite{elshabrawy2018closed}, the symbol error rate (SER) of LoRa with non-coherent detection approximated by:
\begin{equation}\label{eq:ser_lcss_L1}
P_{\rm s} \approx Q\left( \frac{\sqrt{\sfrac{E_{\rm s}}{N_0}} - (H_{M-1}^2 - \sfrac{\pi^2}{12})^\frac{1}{4}}{\sqrt{H_{M-1} - (H_{M-1}^2 - \sfrac{\pi^2}{12})^\frac{1}{2} + \sfrac{1}{2}}}\right),
\end{equation}
where $Q(z) = \sfrac{1}{\sqrt{2\pi}} \int_{z}^{\infty} \exp(-u^2) \mathrm{d}u$ is the $Q$-function and $H_m = \sum_{i=1}^m \sfrac{1}{i}$ is the $m$th harmonic number. Being $\lambda=\log_2(M)$, the conversion from SER to BER of $M$-ary FSK modulation is given by \cite{xiong2006digital}:
\begin{equation}\label{eq:ber_lcss_L1}
P_{\rm b} = \frac{2^\lambda-1}{M-1}P_{\rm s}.
\end{equation}

For $L>1$, the chirps with different slopes interfere with each other. For an unitary energy transmission, i.e., $E_{\rm s} = 1$, the interference power on the $l$th layer is referred to as \(I_1\) and can be extracted from eq. (\ref{eq40}). Taking into account the increase in noise due to interference, assuming that the interference is the same for all $l$, we can simply use $I = I_l$. We now can define the energy per symbol per layer over noise and interference as $\sfrac{E_{\rm s}}{(L(N_0+E_{\rm s} I))}$, and plugging it into \eqref{eq:ser_lcss_L1}, and plug it into \eqref{eq:ser_lcss_L1} and \eqref{eq:ber_lcss_L1}, we have the approximate BER for LCSS as:
\begin{equation}\label{eq:ber_lcss}
P_{\rm b} \approx \frac{2^\lambda-1}{M-1}Q\left( \frac{\sqrt{\sfrac{E_{\rm s}}{(L(N_0+E_{\rm s} I))}} - (H_{M-1}^2 - \sfrac{\pi^2}{12})^\frac{1}{4}}{\sqrt{H_{M-1} - (H_{M-1}^2 -\sfrac{\pi^2}{12})^\frac{1}{2} + \sfrac{1}{2}}}\right).
\end{equation}
\subsubsection{Coherent Detection}
For coherent detection, no close form solution for the SER is known.  So, we can resort to the upper bound assuming equiprobable symbols, which is tight for SER smaller than $10^{-3}$ \cite{xiong2006digital} when $L=1$. In general, considering the same assumptions as in the last section, the upper bound for any $L$ is given by:
\begin{equation}\label{eq:ser_ub_lcss}
P_{\rm s} \leq Q\left( \sqrt{\frac{E_{\rm s}}{L(N_0+E_{\rm s} I)}}\right),
\end{equation}
where $I=0$ for $L=1$, recovering the original expression given in \cite{xiong2006digital}. Analogously to \eqref{eq:ber_lcss_L1}, we have:
\begin{equation}\label{eq:ber_ub_lcss}
P_{\rm b} \leq  \frac{2^\lambda-1}{M-1} Q\left( \sqrt{\frac{E_{\rm s}}{L(N_0+E_{\rm s} I)}}\right).
\end{equation}

\subsection{Layered Dual-Mode Chirp Spread Spectrum}
Because LDMCSS performs the frequency shift on even and odd carriers, it is equivalent to LCSS with $M/2$. 
\subsubsection{Non-Coherent Detection}
The BER expression of LDMCSS for non-coherent detection is given as:
\begin{equation}\label{eq:ber_ldmcss}
P_{\rm b} \approx \frac{2^{\tilde{\lambda}}-1}{\sfrac{M}{2}-1}Q\left( \frac{\sqrt{\sfrac{E_{\rm s}}{(L(N_0+E_{\rm s} I))}} - (H_{\sfrac{M}{2}-1}^2 - \sfrac{\pi^2}{12})^\frac{1}{4}}{\sqrt{H_{\sfrac{M}{2}-1} - (H_{\sfrac{M}{2}-1}^2 -\sfrac{\pi^2}{12})^\frac{1}{2} + \sfrac{1}{2}}}\right),
\end{equation}
where \(\tilde{\lambda} = \log_2(\sfrac{M}{2})\).
\subsubsection{Coherent Detection}
Similarly, the  upper bound for the BER expression of coherent detection of LDMCSS is given as:
\begin{equation}\label{eq:ber_ub_ldmcss}
P_{\rm b} \leq  \frac{2^{\tilde{\lambda}}-1}{\sfrac{M}{2}-1} Q\left( \sqrt{\frac{E_{\rm s}}{L(N_0+E_{\rm s} I)}}\right).
\end{equation}

\section{Simulation Results and Discussions}
In this section, we provide the performances of LCSS and LDMCSS and compare them with state-of-the-art schemes. It is recalled that both LCSS and LDMCSS permit both coherent and non-coherent detection; therefore, we shall illustrate their performances with different detection mechanisms separately. One of the primary characteristics of the schemes proposed is their SE; accordingly, the benchmarking methods that we consider are also capable of achieving high spectral efficiencies. Hence, we consider TDM-CSS, IQ-TDM-CSS, and DM-TDM-CSS. Apart from IQ-TDM-CSS, all the above schemes permit coherent and non-coherent detection. We also consider LoRa for benchmarking as it is a widely adopted scheme.

The performances of proposed schemes are compared in terms of (i) SE; (ii) SE vs. EE performance; (iii) BER performance in AWGN channel; (iv) BER performance in frequency-selective fading channel; (v) BER performance considering frequency offsets (FOs); and (vi) BER performance considering phase offsets (POs).
\subsection{Spectral Efficiency}
Unlike the classical CSS schemes, which provide flexibility in adjusting the SE only by altering \(M\), the proposed schemes are adaptive spectral efficiencies by tailoring \(M\), and \(L\) or \(\tilde{L}\). Consider that a LoRa transmits \(\lambda= \log_2(M)\) bits per symbol. In LCSS, the symbol at each layer is equivalent to the LoRa symbol that transmits \(\lambda\) bits. Therefore, we have \(L\lambda\) transmitted bits when we have \(L\) layers for LCSS. For a particular case, when we have \(L= 2\), the SE and other properties of LCSS are equivalent to the TDM-CSS scheme. On the other hand, in each LDMCSS symbol, even FS at each layer transmits \(\lambda-1=\log_2(\sfrac{M}{2})\) bits; the same is the case with the odd FSs, which also transmit \(\lambda-1\) bits; therefore, a total of \(2\lambda-2\) bits are transmitted per layer. Hence, LDMCSS transmits \(\tilde{L}(2\lambda-2)\) bits per symbol for \(\tilde{L}\) layers. Moreover, for \(\tilde{L} = 2\), LDMCSS is equivalent to DM-TDM-CSS, which transmits \(4\lambda-4\) bits per symbol, possibly changing the SE only with a change in \(M\). The spectral efficiencies of LCSS and LDMCSS schemes, along with the classical counterparts, are listed in the Table \ref{tab1}. IQ-TDM-CSS effectively multiplexes two symbols, each transmitting \(\lambda\) bits in the in-phase component and another \(\lambda\) bits in the quadrature component. Thus, a total of \(4\lambda\) bits per IQ-TDM-CSS symbol.

It is evident that the proposed schemes provide higher SE than other alternatives and the flexibility of tailoring the SE by changing \(L\) or \(\tilde{L}\) and \(M\). Such flexibility does not exist for other alternative schemes.
\begingroup
\setlength{\tabcolsep}{6pt} 
\renewcommand{\arraystretch}{1.2} 
\begin{table}[tb]
\caption{Spectral efficiencies of different CSS schemes}
\centering
\begin{tabular}{c||c}
\hline
\hline
{\textbf{CSS Scheme}} & {\textbf{SE, \(\eta\)  (bits/s/Hz)}}\\
\hline
\hline
{LoRa}& {\(\frac{\lambda}{M}\)}\\
\hline
{TDM-CSS}&{\(\frac{2\lambda}{M}\)}\\
\hline
{IQ-TDM-CSS}&{\(\frac{4\lambda}{M}\)}\\
\hline
{DM-TDM-CSS}&{\(\frac{4\lambda-4}{M}\)}\\
\hline
{LCSS}&{\(\frac{L\lambda}{M}\)}\\
\hline
{LDMCSS}&{\(\frac{\tilde{L}(2\lambda-2)}{M}\)} \\
\hline 
\hline
\end{tabular}
\label{tab1}
\end{table}
\endgroup
\subsection{Spectral Efficiency vs Energy Efficiency Analysis}
%

%
\begin{figure}[t]\centering
\includegraphics[trim={5 0 0 0},clip,scale=4.6]{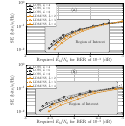}
  \caption{SE versus EE performance for LCSS and LDMCSS considering coherent (a) and non-coherent (b) detection mechanisms for target BER of \(10^{-3}\) in AWGN channel.}
\label{fig7}
\end{figure}
This section evaluates and compares the SE versus EE performance of the proposed schemes with other state-of-the-art techniques. To assess the performance at a given SE, we attain the EE by evaluating \(E_\mathrm{b}/N_0=\frac{E_\mathrm{s}}{\eta N_0}\) required to reach a BER of \(10^{-3}\). Here \(\eta\) refers to the SE of any given scheme and changes by varying \(\lambda = \llbracket 7,12\rrbracket\). Table \ref{tab1} provides the relationship between \(\eta\) and \(\lambda\). The performance is evaluated for all the considered methods in an AWGN channel. We first illustrate the performance of the proposed schemes by changing the number of layers and then compare their performance with the classical counterparts. 

Fig. \ref{fig7}(a) and (b) illustrate the SE versus EE performance of the proposed schemes by modifying \(L\) or \(\tilde{L}\) and utilizing coherent and non-coherent detection mechanisms, respectively. Interestingly, we observe that with an increase in \(L\) or \(\tilde{L}\), the SE increases considerably without imposing a significant penalty on the required \(E_\mathrm{b}/N_0\) for the given BER. The increase in needed \(E_\mathrm{b}/N_0\) is because more layers affects the sparsity of the un-chirped symbol. Since the un-chirped symbol would be more sparse for lesser number of layer, the required \(E_\mathrm{b}/N_0\) would also be less. We also observe another trend, i.e., the achievable SE of LCSS is higher than that of LDMCSS. Moreover, for both LCSS and LDMCSS, the interference for low \(M\) is higher. Therefore, the EE of both schemes is better for higher values of \(M\). Thus, we define a region of interest in which the interference is lower. This region of interest includes \(\lambda = \llbracket 8,12\rrbracket\); however, it shrinks with an increase in \(L\) or \(\tilde{L}\). Overall, for different layers, the performance of LCSS in terms of both EE and SE is marginally better than that of LDMCSS. We also gather from Fig. \ref{fig7} that generally the coherent detector performs relatively better than the non-coherent detection in the AWGN channel. 
\begin{figure}[t]\centering
\includegraphics[trim={5 0 0 0},clip,scale=4.6]{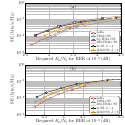}
  \caption{SE versus EE performance for LCSS (with \(L=4\)) and LDMCSS (with \(\tilde{L}=2\)) with other state-of-the-art schemes considering coherent (a) and non-coherent (b) detection mechanisms for target BER of \(10^{-3}\) in AWGN channel.}
\label{fig8}
\end{figure}

\begin{figure}[t]\centering
\includegraphics[trim={5 0 0 0},clip,scale=4.6]{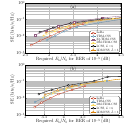}
  \caption{SE versus EE performance for LCSS (with \(L=6\)) and LDMCSS (with \(\tilde{L}=3\)) with other state-of-the-art schemes considering coherent (a) and non-coherent (b) detection mechanisms for target BER of \(10^{-3}\) in AWGN channel. }
\label{fig9}
\end{figure}

\begin{figure}[t]\centering
\includegraphics[trim={5 0 0 0},clip,scale=4.6]{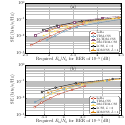}
  \caption{SE versus EE performance for LCSS (with \(L=8\)) and LDMCSS (with \(\tilde{L}=4\)) with other state-of-the-art schemes considering coherent (a) and non-coherent (b) detection mechanisms for target BER of \(10^{-3}\) in AWGN channel.}
\label{fig10}
\end{figure}

Next, we compare the performance of the LCSS and LDMCSS by varying the number of layers with other counterparts. For LCSS, we consider \(L = \{4,6,8\}\), and for LDMCSS, we consider \(\tilde{L} = \{2,3,4\}\). The result for \(\{L,\tilde{L}\} = \{4,2\}\), \(\{L,\tilde{L}\} = \{6,3\}\) and \(\{L,\tilde{L}\} = \{8,4\}\) are illustrated in Fig. \ref{fig8}, Fig. \ref{fig9}, and Fig. \ref{fig10}, respectively. 

The SE of LCSS is the same as that of IQ-TDM-CSS when \(L = 4\). Moreover, the SE of DM-TDM-CSS is the same as that of LDMCSS with \(L= 2\). From Fig. \ref{fig8} (a) and (b), considering coherent detection, IQ-TDM-CSS performs marginally better than LCSS for low values of \(\lambda\) (cf. Fig. \ref{fig8}(a)). On the other hand, the LCSS performs best for non-coherent detection. It is highlighted that IQ-TDM-CSS cannot be detected non-coherently. The performances LCSS is significantly better than TDM-CSS and LoRa for both coherent and non-coherent detections. We also observe that for \(\tilde{L}=2\), LDMCSS and DM-TDM-CSS have essentially the same signal structure, and so does the SE vs. EE performance considering both coherent and non-coherent detection mechanisms. For coherent detection, LDMCSS also performs better than both TDM-CSS and LoRa; however, for non-coherent detection, the performance of TDM-CSS is also similar to that of LDMCSS and DM-TDM-CSS. 
%
%

Fig. \ref{fig9} illustrates that with an increase in \(L\) and \(\tilde{L}\), the SE increases significantly, and so does the EE. We observe that now, for \(L = 6\), the performance of LCSS is markedly better than other counterparts for both the coherent and non-coherent detection (cf. Fig. \ref{fig9}(a) and (b)). On the other hand, considering coherent detection and \(\tilde{L}=3\), we follow that LDMCSS performs better than DM-CSS, TDM-CSS, and LoRa; however, its performance is marginally worse than IQ-TDM-CSS. For non-coherent detection, LDMCSS performs better than all the other counterparts. 

Fig \ref{fig10} illustrates the SE vs. EE performance considering \(L = 8\) for LCSS and \(\tilde{L}= 4\) for LDMCSS. We observe that for both coherent and non-coherent detection, the performance of LCSS in terms of achievable SE and EE is superior to the other schemes. LDMCSS, on the other hand, performs marginally worse than LCSS for non-coherent detection, whereas, for coherent detection, IQ-TDM-CSS is marginally better than LDMCSS. It is reiterated that, unlike LDMCSS, IQ-TDM-CSS has some limitations, such as no possibility of non-coherent detection and high sensitivity to the phase and frequency offsets. The latter feature shall become evident in the sequel of the article.  
\subsection{BER Performance in AWGN Channel}
This section compares the BER performance of LCSS and LDMCSS for \(L = \{4, 6, 8\}\), and \(\tilde{L} = \{2, 3, 6\}\). The results are then compared to other methods assuming \(M = 1024\). Note that when \(L=4\), LCSS has a SE of \(0.3906\) bits/s/Hz. For \(L=6\), the SE is \(0.05859\) bits/s/Hz, and \(L=8\) results in an SE of \(0.07813\) bits/s/Hz. For LDMCSS, \(\tilde{L}=\{2, 3, 4\}\) has SEs of \(0.03516\) bits/s/Hz, \(0.05273\) bits/s/Hz, and \(0.07031\) bits/s/Hz.
%
\begin{figure}[t]\centering
\includegraphics[trim={10 0 0 0},clip,scale=2.32]{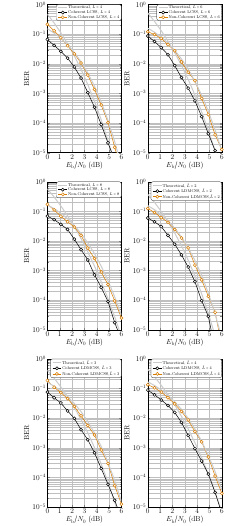}
  \caption{BER performance comparison of LCSS and LDMCSS considering coherent (a)  and non-coherent (b) detection mechanisms in AWGN channel for \(M=1024\).}
\label{fig11}
\end{figure}

Fig. \ref{fig11} compares the BER performance of LCSS and LDMCSS for different numbers of layers under both coherent and non-coherent detection. The numerical simulation results are compared with the closed-form BER expressions for the proposed schemes obtained for non-coherent detection. Moreover, for coherent detection, we provide the theoretical upper-bound on the BER for both schemes. It can be observed from Fig. \ref{fig11}, that the numerical results for the non-coherent detection match the theoretical results, whereas, for coherent detection the upper-bound of the BER expressions is tight when compared with the numerical simulations.  It also demonstrates that LCSS have a slightly better performance. As the number of layers increases, the SE increases but the BER also increases slightly. For instance, when \(L\) increases from \(4\) to \(6\) for LCSS, the SE increases by \(50\%\), while the \(E_\mathrm{b}/N_0\) required for a BER of \(10^{-3}\) increases by less than \(0.4\) dB for both coherent and non-coherent detection. Similar trends are seen for LDMCSS for both coherent and non-coherent detection mechanisms. 

\begin{figure}[t]\centering
\includegraphics[trim={10 0 0 0},clip,scale=2.4]{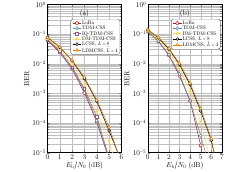}
  \caption{BER performance comparison of LCSS (with \(L=8\)) and LDMCSS (with \(\tilde{L}=4\)) with other state-of-the-art schemes considering coherent (a) and non-coherent (b) detection mechanisms in AWGN channel.}
\label{fig12}
\end{figure}
Fig. \ref{fig12} compares the BER performance of LCSS and LDMCSS with TDM-CSS, DM-TDM-CSS, IQ-TDM-CSS, and LoRa, using  \(M = 1024\). Fig. \ref{fig12}(a) shows the BER performance under coherent detection, and Fig. \ref{fig12}(b) illustrates the BER performance under non-coherent detection. For the given \(M\), the SE of TDM-CSS, DM-TDM-CSS, IQ-TDM-CSS, and LoRa is \(0.01953\) bits/s/Hz, \(0.03516\) bits/s/Hz, \(0.03906\) bits/s/Hz, and \(0.0097\) bits/s/Hz respectively. In the comparison, LCSS with \(L = 8\) and LDMCSS with \(\tilde{L} = 4\) are considered, which results in SE of \(0.0781\) bits/s/Hz and \(0.0703\) bits/s/Hz, respectively. 

Figs. \ref{fig12}(a) and (b) show that LoRa and TDM-CSS have the best BER performance under coherent detection. IQ-TDM-CSS has slightly worse performance compared to LoRa and TDM-CSS. LCSS and LDMCSS have similar BER performance, which is higher than the other methods; however, it is worth noting that LCSS is \(2\) times more spectrally efficient than IQ-TDM-CSS, \(2.22\) times more than DM-TDM-CSS, \(4\) times more than TDM-CSS, and \(8\) times more than LoRa. Similarly, LDMCSS provides \(1.8\) times, \(2\) times, \(3.6\) times, and \(7.2\) times more SE than IQ-TDM-CSS, TDM-CSS, DM-TDM-CSS, and LoRa, respectively. Therefore, considering the improvement in SE, the increase in BER is only minimal. For example, to achieve the same BER under coherent detection, LCSS and LDMCSS require \(0.5\) dB, \(0.7\) dB, \(0.4\) dB, and \(0.8\) dB higher \(E_\mathrm{b}/N_0\) than IQ-TDM-CSS, TDM-CSS, DM-TDM-CSS, and LoRa, respectively. Under non-coherent detection, LCSS needs \(0.4\) dB, \(0.2\) dB, and \(0.4\) dB higher \(E_\mathrm{b}/N_0\) than TDM-CSS, DM-TDM-CSS, and LoRa, respectively. In contrast, non-coherent LDMCSS needs \(0.42\) dB, \(0.22\) dB, and \(0.42\) dB higher \(E_\mathrm{b}/N_0\) than TDM-CSS, DM-TDM-CSS, and LoRa. Therefore, even though LCSS and LDMCSS have slightly higher BER than other methods, the difference is relatively small, considering the improvements in SE they provide.
\subsection{BER Performance in Fading Channel}
\begin{figure}[t]\centering
\includegraphics[trim={10 0 0 0},clip,scale=2.6]{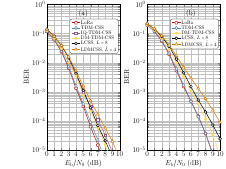}
  \caption{BER performance comparison of LCSS (with \(L=8\)) and LDMCSS (with \(\tilde{L}=4\)) with other state-of-the-art schemes considering coherent (a) and non-coherent (b) detection mechanisms in a frequency-selective fading channel.}
\label{fig13}
\end{figure}

This section examines the performance of LCSS and LDMCSS over a frequency-selective \(2\)-tap fading channel with impulse response of \(h(n) = \sqrt{1-\rho}\delta (nT) + \sqrt{\rho}\delta(nT-T)\), where \(T\) is the sampling duration, and \(0\leq \rho \leq 1\). The results are presented in Fig. \ref{fig13} for both coherent and non-coherent detection with a specific value of \(\rho = 0.2\). 
%

When analyzing the BER performance of LCSS and LDMCSS under coherent detection, as shown in Fig. \ref{fig13}(a), we can see that LCSS performs better than IQ-TDM-CSS and LDMCSS. Additionally, the BER performance of LCSS indicates only a slight degradation compared to DM-TDM-CSS. LoRa has the best BER performance among all the methods. It is also notable that LDMCSS has the worst BER performance, primarily due to the signal structure, making it more susceptible to the frequency selective fading effects.

When analyzing the BER performance of LCSS and LDMCSS under non-coherent detection, as shown in Fig. \ref{fig13}(b), it can be seen that there is a slight increase in the BER of LCSS compared to DM-TDM-CSS. Moreover, the difference in the BER performance between LCSS and LDMCSS is now more noticeable.

Although LCSS and LDMCSS may not have the best BER performance, their higher SE and only slight deterioration make them noteworthy.
\subsection{BER Performance considering Frequency Offset}
In this section, we examine the effect of carrier frequency offset (FO) on BER performance. The FO causes a linear accumulation of phase rotations from one symbol to another. The equation represents the received symbol, including the FO impact is:
\begin{equation}\label{eq57}
y(n) = \exp\left\{\frac{j2\pi \Delta f n}{M}\right\} s(n) + w(n), 
\end{equation}
where \(\Delta f\) is the FO normalized to the frequency spacing and could be seen as residual FO because IoT modems, like Bluetooth usually have a carrier frequency offset compensator before demodulation; therefore, after compensation, \(\Delta f\) reflects the residual FO. To evaluate BER performance, we consider \(\Delta f\) of \(0.2\) and \(M=1024\), and AWGN channel. Moreover, we employ both the coherent and non-coherent detection mechanisms. 
\begin{figure}[t]\centering
\includegraphics[trim={10 0 0 0},clip,scale=2.6]{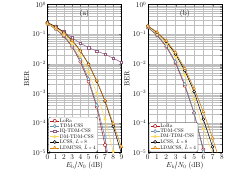}
  \caption{BER performance comparison of LCSS (with \(L=8\)) and LDMCSS (with \(\tilde{L}=4\)) with other state-of-the-art schemes considering coherent (a) and non-coherent (b) detection mechanisms with a FO of \(\Delta f=0.2\) Hz.}
\label{fig14}
\end{figure}

Fig. \ref{fig14} shows the BER performance in the presence of FO for both coherent detection (Fig. \ref{fig14}(a)) and non-coherent detection (Fig. \ref{fig14}(b)). It can be seen that LCSS and LDMCSS have similar performance when using coherent detection. IQ-TDM-CSS has the worst performance due to its sensitivity to FO. The other schemes, such as LoRa and TDM-CSS have the best performance. Similarly, DM-TDM-CSS shows a slight improvement over LCSS and LDMCSS. For non-coherent detection, LoRa and TDM-CSS have similar performance. However, non-coherently detected LCSS performs better than non-coherently detected LDMCSS. The performance of DM-TDM-CSS falls between that of LCSS and LoRa/TDM-CSS. Another discernible trend is that the non-coherent detector performs relatively better than the coherent detector in the presence of frequency offsets. 
\subsection{BER Performance considering Phase Offset}
This section examines the performance of all the schemes in the presence of phase offset (PO) commonly found in low-cost devices. The received symbol corrupted by AWGN in the presence of PO is represented by the equation:
\begin{equation}\label{eq58}
y(n) = \exp\{j\psi\} s(n) + w(n),
\end{equation}
where \(\psi\) is the PO. We evaluate the performance of the considered schemes under both coherent and non-coherent detection, using a PO of  $\psi = \sfrac{\pi}{4}$ radians and $M = 1024$.

\begin{figure}[t]\centering
\includegraphics[trim={10 0 0 0},clip,scale=2.6]{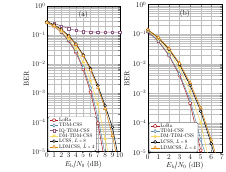}
  \caption{BER performance comparison of LCSS (with \(L=8\)) and LDMCSS (with \(\tilde{L}=4\)) with other state-of-the-art schemes considering coherent (a) and non-coherent (b) detection mechanisms with a PO of $\psi = \sfrac{\pi}{4}$ radians.}
\label{fig15}
\end{figure}

The BER performance of proposed schemes in the presence of POs is compared in Fig. \ref{fig15}(a) and (b) using both coherent and non-coherent detection, respectively. It is important to note that IQ-TDM-CSS is highly sensitive to POs as it carries information in both in-phase and quadrature components that drastically affect the integrity of the symbols. The results show that non-coherent detection performs better than coherent detection when POs are present. In coherent detection, LDMCSS performs slightly better than LCSS, requiring \(0.1\) dB less \(E_\mathrm{b}/N_0\) to reach a BER of \(10^{-3}\). However, this is reversed when using coherent detection, where LCSS requires \(0.1\) dB less \(E_\mathrm{b}/N_0\) than LDMCSS. For coherent detection, LCSS needs \(1.3\) dB, \(1.15\) dB, and \(0.8\) dB higher \(E_\mathrm{b}/N_0\) than LoRa, TDM-CSS, and DM-TDM-CSS, respectively, to reach a BER of \(10^-3\). Meanwhile, LDMCSS requires \(1.2\) dB, \(1.05\) dB, and \(0.7\) dB higher \(E_\mathrm{b}/N_0\) than LoRa, TDM-CSS, and DM-TDM-CSS, respectively, for the same target BER. When using non-coherent detection, LCSS requires \(0.7\) dB, \(0.6\) dB, and \(0.3\) dB higher \(E_\mathrm{b}/N_0\) than LoRa, TDM-CSS, and DM-TDM-CSS, respectively, indicating that LCSS performs better than other schemes when non-coherent detection is used. Similarly, for non-coherent detection, LDMCSS requires \(0.1\) dB higher \(E_\mathrm{b}/N_0\) than LCSS over other methods.
%
%
\begin{figure}[tb]\centering
\includegraphics[trim={26 0 0 0},clip,scale=0.87]{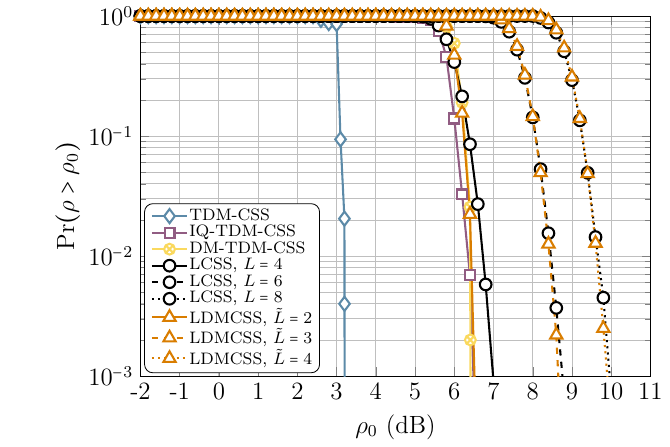}
  \caption{PAPR CCDFs for different schemes analyzed in this work. The CCDF curves are obtained for \(\lambda =8\).}
\label{ccdf}
\end{figure}
\subsection{Merits and Limitations of Proposed Schemes}
In this section, we first indicate some improvements that the proposed schemes offer relative to other counterparts. Then, we also list some of the limitations of the proposed LCSS and LDMCSS schemes. Some noticeable improvements of LCSS and LDMCSS are:
\begin{enumerate}
\item The proposed schemes offer significant improvements in \textit{flexibility and a wide range of spectral efficiencies} compared to other counterparts. The proposed schemes offer an additional degrees of freedom through the number of layers, which enables a wider range of achievable SEs. In contrast, most state-of-the-art schemes are limited to a lower SE, which can be tuned only through \(M\).
\item The proposed schemes offer a \textit{higher SE} than classical schemes. 
\item The proposed schemes offer the flexibility of utilizing both \textit{coherent and non-coherent detection mechanisms}, which is more suitable for low-cost devices commonly used in LPWANs.  While some traditional schemes can also use both detection mechanisms, they do not offer the same level of SE as the proposed schemes, or they cannot use either of the detection mechanisms altogether. The ability to use non-coherent detection can be a key advantage for cost-effective devices with power constraints.
\item The proposed schemes have a \textit{broad range of applicability} due to their adaptability to various types of communication. Their flexibility in terms of system efficiency allows them to be tailored to specific applications, considering factors such as data rate, range, and interference environment, making them more suitable than traditional methods.
\end{enumerate}

On the other hand, some discernible limitations of the proposed schemes are:
\begin{enumerate}
\item The proposed schemes do not have a \textit{constant envelope}, which can present challenges during their practical implementation. This would increase the peak-to-average power ratio (PAPR), making it susceptible to non-linear distortions in the system. However, it is worth noting that many other CSS schemes capable of achieving higher SE than the classical LoRa, also do not have a constant envelope. This is because these schemes either use PSs or additional FSs to attain higher SE. PAPR, \(\rho\) is mathematically defined as:
\begin{equation}
\rho = \frac{\max \left\{\vert s(n)\vert^2\right\}}{\frac{1}{M}\sum_{n=0}^{M-1}\vert s(n)\vert^2}.
\end{equation}

To graphically illustrate PAPR, the complementary cumulative distribution function (CCDF) is used, which measures the probability that \(\rho\) exceeds a specified threshold, \(\rho_0\), i.e., \(\mathrm{Pr}(\rho>\rho_0)\). Fig. \ref{ccdf} depicts the CCDFs for PAPR of various schemes considered in this study, all evaluated for \(\lambda=8\). For LCSS, we consider \(L\) equal to \(4,6\) and \(8\), whereas for LDMCSS, we consider \(\tilde{L} = 2,3\) and \(4\). It can be observed that with an increase in the number of layers for both LCSS and LDMCSS, the PAPR also increases. However, we also notice that the other counterparts, which activate multiple FS, also have significantly higher PAPR. The results in Fig. \ref{ccdf} indicate that a trade-off is necessary regarding how many layers should be used, as the higher the number of layers, the higher will be the non-linear distortions. However, based on the component's non-linear characteristics, the number of layers for the proposed schemes can be tuned to avoid these non-linear distortions. This flexibility is not possible with other existing schemes in the literature. 

The transceiver, especially the receiver for the proposed schemes, is more \textit{complex} than other CSS-based counterparts, requiring more arithmetic operations to demodulate the signal. It can be noticed that the number of necessary DFTs to retrieve information from the received symbols is equal to \(L\) and \(\tilde{L}\) for LCSS and LDMCSS, respectively. Thus, increasing the number of layers would also increase the receiver complexity. On the other hand, DM-TDM-CSS, IQ-TDM-CSS, and TDM-CSS require the computation of two DFTs to demodulate the received symbol. However, it is important to notice that the increase in complexity is not significant compared to the improvement in the spectral and energy efficiency, e.g., LCSS with \(L= 6\) and LDMCSS with \(\tilde{L} = 3\) exhibit a spectral efficiency that is six times that of classical LoRa. However, LCSS requires five additional computations of DFTs, and LDMCSS requires two additional computations of DFTs. Thus, it is apparent that the proposed schemes reduce over-the-air transmission time compared to the classical counterparts. Assuming DFT is implemented via fast Fourier transform (FFT), the arithmetic operations needed for the studied schemes are summarized in Table \ref{tab_complex}. This increase in receiver complexity would have minimal impact on the longevity of the battery-powered terminals, as it primarily depends on the transmit power. Since we have comprehensively established that the proposed schemes are more power efficient than other counterparts, we believe that the transmit power required would be significantly less, improving the longevity of the battery-powered devices in LPWANs. On the other hand, as far as an increase in power consumption related to an increase in receiver complexity is concerned, that would be minimal because the power consumption of computing additional FFT would be significantly less. Moreover, employing the proposed schemes, the battery-powered terminals would have to transmit less frequently, minimizing power consumption.
\begingroup
\setlength{\tabcolsep}{6pt} 
\renewcommand{\arraystretch}{1.1} 
\begin{table}[tb]
\caption{Receiver complexities of different CSS schemes.}
\centering
\begin{tabular}{c||c}
\hline
\hline
\textbf{CSS Scheme} & \textbf{Arithmetic Operations}\\
\hline
\hline
LoRa& \(4M\log_2(M)-6M+8\)\\
\hline
TDM-CSS&\(8M\log_2(M)-12M+16\)\\
\hline
IQ-TDM-CSS&\(8M\log_2(M)-12M+16\)\\
\hline
DM-TDM-CSS&\(8M\log_2(M)-12M+16\)\\
\hline
LCSS&\(4LM\log_2(M)-6LM+8L\)\\
\hline
LDMCSS&\(4\tilde{L}M\log_2(M)-6\tilde{L}M+8\tilde{L}\) \\
\hline 
\hline
\end{tabular}
\label{tab_complex}
\end{table}
\endgroup
%
\item As \(L\) increases, then so does the complexity and interference level of the LCSS and LDMCSS schemes. Because of this increase in interference, the energy efficiency starts to decrease, reducing the region of interest in SE for the proposed schemes.
\end{enumerate}
\subsection{Selecting Number of Layers}
Selecting the number of layers for the proposed LCSS and LDMCSS is a critical aspect. On the one hand, increasing the number of layers increases the SE and EE; however, on the other hand, it also increases interference power, complexity, and PAPR. The improvement in SE and EE increases the transmission distances, and more information can be transmitted; therefore, the transmitter has to communicate less frequently, reducing power consumption. However, this also restricts using higher SF because of the increase in interference power for lower \(M\) (i.e., lower SF). Moreover, the non-linear distortion may also increase due to higher PAPR. Selecting the number of layers depends primarily on the application requirements and the characteristics of the equipment in use. If the components have a narrow linear range, using less number of layers may be preferable to avoid non-linear distortions. Also, fewer layers can be used if lower data rates are needed at shorter distances. Considering all these characteristics, we believe using \(L = \{6,8\}\) and \(\tilde{L} = \{3,4\}\) could provide a good compromise between an increase in SE, improvement in EE, and the impact of an increase in interference power and non-linear distortions. Using these much number of layers improves the SE and EE significantly while not proving so harsh on the interference power and non-linear distortions.

\section{Conclusions}
The article introduces two novel waveform designs, LCSS and LDMCSS, as alternatives to the LPWAN technology, LoRa. These designs proffer greater adaptability in adjusting data rates for various applications and conditions and possess high achievable SE and EE compared to the current state-of-the-art options. Despite their similarities, LCSS and LDMCSS differ in the intricacies of their waveform design, with LDMCSS necessitating fewer layers, thus rendering it less complex. However, it must be noted that the layered structure of these designs can lead to interference among layers, particularly when lower alphabet cardinalities and more layers are involved. Despite certain limitations, such as the high receiver complexity, interference due to the layered structure, and the PAPR, the proposed schemes have exhibited a higher SE than other LPWAN alternatives. Furthermore, these schemes can employ both coherent and non-coherent detection mechanisms and are relatively robust against frequency and phase offsets that may occur in low-cost devices. While LCSS demonstrates slightly better performance than LDMCSS, it is computationally complex. We hope that these new CSS-based designs, presented herein, inspire further research aimed at enhancing their performance.

\bibliographystyle{unsrt}
\bibliography{biblio}

\appendices
\section{Orthogonality and Interference Analysis of Benchmark Schemes}
It is highlighted that, like the proposed schemes, the benchmark schemes considered in this work, TDM-CSS, IQ-TDM-CSS, and DM-TDM-CSS, are also not orthogonal and result in interference due to time-domain multiplexing. This appendix briefly provides the orthogonality and interference analysis results for these benchmark schemes.
 \subsection{TDM-CSS}\label{first_ap}
The transmit symbol for TDM-CSS is given as:
\begin{equation}\label{ap_eq1}
\begin{split}
s(n)= \exp\left\{j\frac{\pi}{M}\left(2k_1n+n^2\right)\right\} +\exp\left\{j\frac{\pi}{M}\left(2k_2n-n^2\right)\right\}.
\end{split}
\end{equation}

To evaluate the orthogonality of TDM-CSS symbols, we consider the inner product of \(\boldsymbol{s}\) and \(\overline{\boldsymbol{s}}\), i.e., \(\langle \boldsymbol{s},\overline{\boldsymbol{s}}\rangle\) where the activated FSs are \(\overline{k}_1\) and \(\overline{k}_2\). Assuming that \(k_1 \neq \overline{k}_1\) and \(k_2 \neq \overline{k}_2\),  \(\langle \boldsymbol{s},\overline{\boldsymbol{s}}\rangle\) yields:
\begin{equation}\label{ap_eq2}
\begin{split}
\langle \boldsymbol{s},\overline{\boldsymbol{s}}\rangle &= \alpha \exp\left\{-j\frac{\pi}{2M}\left(k_1-\overline{k}_2\right)^2\right\} \\&~+ \alpha^\ast \exp\left\{j\frac{\pi}{2M}\left(k_2-\overline{k}_1\right)^2\right\},
\end{split}
\end{equation}
where \(\alpha = \sqrt{\sfrac{M}{2}}\exp\left\{j\sfrac{\pi}{4}\right\}\). It can be observed that TDM-CSS symbols are not orthogonal because as activated FSs of the two multiplexed symbols results interference.

As TDM-CSS multiplexes symbols with two different chirp rates; therefore, the received symbol, \(r(n)\), is de-chirped by the down-chirp, \(c_\mathrm{d}(n) = \exp\left\{-j\frac{\pi}{M}n^2\right\}\), and the up-chirp, \(c_\mathrm{u}(n)= c^\ast_\mathrm{u}(n)\), resulting in \(r_1(n) = r(n)c_\mathrm{d}(n)\) and \(r_2(n) = r(n)c_\mathrm{u}(n)\), respectively. Considering that \(\tilde{k}_1= k_1-k\) and \(\tilde{k}_2= k_2-k\), then the \(M\)-order DFT of these de-chirped symbols, \(R_1(k)\) and \(R_2(k)\), are given as:
\begin{equation}\label{ap_eq3}
R_1(k) = \underbrace{M}_{\text{Signal}} + \underbrace{\alpha^\ast \exp\left\{j\frac{\pi}{2M}\tilde{k}_2^2\right\}\beta_1}_{\text{Interference}} + \underbrace{W_1(k)}_{\text{Noise}},
\end{equation}
and
\begin{equation}\label{ap_eq4}
R_2(k) = \underbrace{M}_{\text{Signal}} + \underbrace{\alpha \exp\left\{-j\frac{\pi}{2M}\tilde{k}_1^2\right\}\beta_2}_{\text{Interference}} + \underbrace{W_2(k)}_{\text{Noise}},
\end{equation}
respectively, where
\begin{equation}\label{ap_eq5}
\begin{split}
\beta_1&=1+\exp\left\{j\frac{\pi}{2}\left(M+2\tilde{k}_2\right)\right\}\\&=  \begin{dcases}
    1 & \quad \mathrm{rem}\left(\vert \tilde{k}_2\vert,2\right)=0\\
   -1 & \quad \mathrm{rem}\left(\vert \tilde{k}_2\vert,2\right)=1 \\
  \end{dcases},\\
\beta_2 &=1+\exp\left\{-j\frac{\pi}{2}\left(M+2\tilde{k}_1\right)\right\}\\&=  \begin{dcases}
    1 & \quad \mathrm{rem}\left(\vert \tilde{k}_1\vert,2\right)=0\\
   -1 & \quad \mathrm{rem}\left(\vert \tilde{k}_1\vert,2\right)=1 \\
  \end{dcases}.
\end{split}
\end{equation}

Moreover, \(W_1(k)\) and \(W_2(k)\) are the DFT outputs for \(w_1(n) = w(n)c_\mathrm{d}(n)\) and  \(w_2(n) = w(n)c_\mathrm{u}(n)\), respectively. From (\ref{ap_eq3}) and (\ref{ap_eq4}), we gather that the activated FS of the two multiplexed symbols causes interference with each other.

 \subsection{IQ-TDM-CSS}\label{second_ap}
The transmit symbol for IQ-TDM-CSS is given as:
\begin{equation}\label{ap_eq6}
\begin{split}
s(n) &= \exp\left\{j\frac{\pi}{M}\left(2k_i n+n^2\right)\right\} + j\exp\left\{j\frac{\pi}{M}\left(2k_q n+n^2\right)\right\}\\&+\exp\left\{j\frac{\pi}{M}\left(2\tilde{k}_i n-n^2\right)\right\}\!+\!j\exp\left\{j\frac{\pi}{M}\left(2\tilde{k}_q n-n^2\right)\right\}.
\end{split}
\end{equation}

The inner product of two IQ-TDM-CSS symbols, \(\boldsymbol{s}\), and \(\overline{\boldsymbol{s}}\), leads to:
\begin{equation}\label{ap_eq7}
\begin{split}
\langle \boldsymbol{s},\overline{\boldsymbol{s}}  \rangle &= \alpha \left(\exp\left\{-j\frac{\pi}{2M}\left(k_i-\tilde{\overline{k}}_i\right)^2\right\}\right.\\
&\left.~ -j\exp\left\{-j\frac{\pi}{2M}\left(k_i-\tilde{\overline{k}}_q\right)^2\right\}\right.\\&
\left.~+j\exp\left\{-j\frac{\pi}{2M}\left(k_q-\tilde{\overline{k}}_i\right)^2\right\}\right.\\&\left.~+ \exp\left\{-j\frac{\pi}{2M}\left(k_q-\tilde{\overline{k}}_q\right)^2\right\}\right)\\
&~+\alpha^\ast \left(\exp\left\{j\frac{\pi}{2M}\left(\tilde{k}_i-\overline{k}_i\right)^2\right\}\right.\\
&\left.~-j\exp\left\{j\frac{\pi}{2M}\left(\tilde{k}_i-\overline{k}_q\right)^2\right\}\right.\\&\left.~+j\exp\left\{j\frac{\pi}{2M}\left(\tilde{k}_q-\overline{k}_i\right)^2\right\} \right.\\&\left.~+ \exp\left\{j\frac{\pi}{2M}\left(\tilde{k}_q-\overline{k}_q\right)^2\right\}\right).
\end{split}
\end{equation}

To attain (\ref{ap_eq7}), we consider that the activated FSs in \(\overline{\boldsymbol{s}}\) are \(\overline{k}_i\), \(\overline{k}_q\), \(\tilde{\overline{k}}_i\), and \(\tilde{\overline{k}}_q\). Moreover, it is also assumed that \(\tilde{\overline{k}}_i \neq k_i\),  \(\overline{k}_q \neq k_q\), \(\tilde{\overline{k}}_i\neq \tilde{k}_i\), and \(\tilde{\overline{k}}_q\neq \tilde{k}_q\). It is apparent that the IQ-TDM-CSS symbols are not orthogonal because the in-phase activated FS of one IQ-TDM-CSS symbol causes interference within both the in-phase and quadrature components of the other IQ-TDM-CSS symbol. Similarly, the activated FS for the quadrature component of one IQ-TDM-CSS symbol causes interference within both the in-phase and quadrature components of the other IQ-TDM-CSS symbol. 

In IQ-TDM-CSS, we again have two multiplexed symbols, both having activated FSs in the in-phase and quadrature components. Like TDM-CSS, the received symbol, \(r(n)\), would be de-chirped using \(c_\mathrm{d}(n)\) and \(c_\mathrm{u}(n)\), as \(r_1(n)= r(n)c_\mathrm{d}(n)\) and \(r_2(n)= r(n)c_\mathrm{u}(n)\). Considering that \(\kappa_i= k_i-k\), \(\kappa_q = k_q-k\), \(\tilde{\kappa}_i= \tilde{k}_i-k\), and \(\tilde{\kappa}_q=\tilde{k}_q-k\), the DFT of \(r_1(n)\) and \(r_2(n)\) results in:
\begin{equation}\label{ap_eq8}
\begin{split}
R_1(k) &= \underbrace{2M}_{\text{Signal}} \\&~+ \underbrace{\alpha^\ast \left(\exp\left\{j\frac{\pi}{2M}\tilde{\kappa}_i^2\right\}\beta_{1,i}+ \exp\left\{j\frac{\pi}{2M}\tilde{\kappa}_q^2\right\}\beta_{1,q}\right)}_{\text{Interference}}\\
&+\underbrace{W_1(k)}_{\text{Noise}},
\end{split}
\end{equation}
and
\begin{equation}\label{ap_eq9}
\begin{split}
R_2(k) &= \underbrace{2M}_{\text{Signal}} \\&~+ \underbrace{\alpha \left(\exp\left\{-j\frac{\pi}{2M}\kappa_i^2\right\}\beta_{2,i}+ \exp\left\{-j\frac{\pi}{2M}\kappa_q^2\right\}\beta_{2,q}\right)}_{\text{Interference}}\\&~+\underbrace{W_2(k)}_{\text{Noise}},
\end{split}
\end{equation}
respectively, where 
\begin{equation}\label{ap_eq9bis}
\beta_{1,(\cdot)} = 1+\exp\left\{j\frac{\pi}{2}\left(M+2\tilde{\kappa}_{(\cdot)}\right)\right\}
\end{equation},
and
\begin{equation}\label{ap_eq9bis2}
\beta_{2,(\cdot)} = 1+\exp\left\{-j\frac{\pi}{2}\left(M+2\kappa_{(\cdot)}\right)\right\},
\end{equation}
where \((\cdot)\) in the subscript is used to denote \(i\) and \(q\). As aforementioned, in IQ-TDM-CSS, two symbols are multiplexed, where each symbol has an activated in-phase and quadrature FS. The results indicate that one symbol's activated in-phase and quadrature FS causes interference with the other. 

 \subsection{DM-TDM-CSS}\label{third_ap}
he transmit symbol for DM-TDM-CSS is given as:
\begin{equation}\label{ap_eq10}
\begin{split}
s(n) &= \exp\left\{j\frac{\pi}{M}\left(2k_{\mathrm{e},1}n\!+\!n^2\right) \right\}\! +\! \exp\left\{j\frac{\pi}{M} \left(2k_{\mathrm{o},1}n\! +\!n^2\right)\right\}\\&~+ \exp\left\{j\frac{\pi}{M}\left(2k_{\mathrm{e},2}n\!-\!n^2\right) \right\}\! +\! \exp\left\{j\frac{\pi}{M} \left(2k_{\mathrm{o},2}n\! -\!n^2\right)\right\}.
\end{split}
\end{equation}

To evaluate the closed-form expressions for orthogonality between two DM-TDM-CSS symbols, we evaluate the inner product of \(\boldsymbol{s}\) and \(\overline{\boldsymbol{s}}\), which yields:

\begin{equation}\label{ap_eq11}
\begin{split}
\langle\boldsymbol{s},\overline{\boldsymbol{s}}\rangle &=\alpha\left(\exp\left\{-j\frac{\pi}{2M}\left(k_{\mathrm{e},1}-\overline{k}_{\mathrm{e},2}\right)^2\right\}\right.\\&\left.~+ \exp\left\{-j\frac{\pi}{2M}\left(k_{\mathrm{o},1}-\overline{k}_{\mathrm{o},2}\right)^2\right\}\right) \\
&~+ \alpha^\ast\left(\exp\left\{j\frac{\pi}{2M}\left(k_{\mathrm{e},2}-\overline{k}_{\mathrm{e},1}\right)^2\right\}\right.\\&\left.~+ \exp\left\{j\frac{\pi}{2M}\left(k_{\mathrm{o},2}-\overline{k}_{\mathrm{o},1}\right)^2\right\}\right).
\end{split}
\end{equation}
It must be noticed that the activated even and odd FSs in \(\overline{\boldsymbol{s}}\) are \(\overline{k}_{\mathrm{e},1}\neq k_{\mathrm{e},1}\), \(\overline{k}_{\mathrm{e},2}\neq k_{\mathrm{e},2}\), \(\overline{k}_{\mathrm{o},1}\neq k_{\mathrm{o},1}\) and \(\overline{k}_{\mathrm{o},2}\neq k_{\mathrm{o},2}\). 

We gather from the symbols structure of the DM-TDM-CSS symbol that two symbols with different chirp rates are multiplexed, where each symbol has an activated even FS and odd FS. Thus, there would be some interference in both the even frequencies and the odd frequencies. Moreover, like TDM-CSS and IQ-TDM-CSS, the received symbol, \(r(n)\), is first de-chirped by multiplying \(c_\mathrm{d}(n)\) and \(c_\mathrm{u}(n)\), which results in \(r_1(n)\) and \(r_2(n)\). The DFT of \(r_1(n)\) and \(r_2(n)\) leads to \(R_1(k)\) and \(R_2(k)\), respectively. The even and odd frequencies of \(R_1(k)\) are respectively given as:
\begin{equation}\label{ap_eq12}
R_1(k_{\mathrm{e},1}) = \underbrace{M}_{\mathrm{Signal}} + \underbrace{2\alpha^\ast\exp\left\{j\frac{\pi}{2M}\tilde{k}_{\mathrm{e},2}^2\right\}}_{\mathrm{Interference}}+W_1(k_{\mathrm{e},1}),
\end{equation}
and
\begin{equation}\label{ap_eq13}
\begin{split}
R_1(k_{\mathrm{o},1}) = \underbrace{M}_{\mathrm{Signal}} + \underbrace{2\alpha^\ast\exp\left\{j\frac{\pi}{2M}\tilde{k}_{\mathrm{o},2}^2\right\}}_{\mathrm{Interference}}+W_1(k_{\mathrm{o},1}),
\end{split}
\end{equation}
respectively, where \(\tilde{k}_{\mathrm{e},2} = {k}_{\mathrm{e},2}-k_{\mathrm{e},1}\) and \(\tilde{k}_{\mathrm{o},2} = {k}_{\mathrm{o},2}-k{\mathrm{o},1}\). Similarly, the even and odd frequencies of \(R_2(k)\) are given as:
\begin{equation}\label{ap_eq14}
R_2(k_{\mathrm{e},2}) = \underbrace{M}_{\mathrm{Signal}} + \underbrace{2\alpha\exp\left\{-j\frac{\pi}{2M}\tilde{k}_{\mathrm{e},1}^2\right\}}_{\mathrm{Interference}}+W_2(k_{\mathrm{e},2}).
\end{equation}
and
\begin{equation}\label{ap_eq15}
R_2(k_{\mathrm{o},2}) = \underbrace{M}_{\mathrm{Signal}} + \underbrace{2\alpha\exp\left\{-j\frac{\pi}{2M}\tilde{k}_{\mathrm{o},1}^2\right\}}_{\mathrm{Interference}}+W_2(k_{\mathrm{o},2}),
\end{equation}
respectively, where \(\tilde{k}_{\mathrm{e},1} = {k}_{\mathrm{e},1}-k_{\mathrm{e},2}\) and \(\tilde{k}_{\mathrm{o},1} = {k}_{\mathrm{o},1}-k_{\mathrm{o},2}\). From (\ref{ap_eq12}) and (\ref{ap_eq13}), we can observe that the activated FSs of the second chirped symbol, i.e., \(k_{\mathrm{e},2}\) and \(k_{\mathrm{o},2}\), cause interference when the activated FSs of the first chirp symbol, i.e., \(k_{\mathrm{e},1}\) and \(k_{\mathrm{o},1}\) are to be determined. We can draw similar conclusions from (\ref{ap_eq14}) and (\ref{ap_eq15}) that the FSs of the first chirped symbol, \(k_{\mathrm{e},1}\) and \(k_{\mathrm{o},1}\), cause interference when we need to determine the FSs of the second chirped symbol, \(k_{\mathrm{e},2}\) and \(k_{\mathrm{o},2}\).

\end{document}